\documentclass[fleqn,usenatbib]{mnras}

\usepackage{newtxtext,newtxmath}
\usepackage[T1]{fontenc}

\DeclareRobustCommand{\VAN}[3]{#2}
\let\VANthebibliography\thebibliography
\def\thebibliography{\DeclareRobustCommand{\VAN}[3]{##3}\VANthebibliography}

\usepackage{graphicx}	% Including figure files
\usepackage{amsmath}	% Advanced maths commands
\newcommand{\aref}[1]{\hyperref[#1]{Appendix~\ref{#1}}}

\usepackage{scalerel,tikz}
\usetikzlibrary{svg.path}
\definecolor{orcidlogocol}{HTML}{A6CE39}
\tikzset{orcidlogo/.pic={
 \fill[orcidlogocol] svg{M256,128c0,70.7-57.3,128-128,128C57.3,256,0,198.7,0,128C0,57.3,57.3,0,128,0C198.7,0,256,57.3,256,128z};
 \fill[white] svg{M86.3,186.2H70.9V79.1h15.4v48.4V186.2z}
 svg{M108.9,79.1h41.6c39.6,0,57,28.3,57,53.6c0,27.5-21.5,53.6-56.8,53.6h-41.8V79.1z M124.3,172.4h24.5c34.9,0,42.9-26.5,42.9-39.7c0-21.5-13.7-39.7-43.7-39.7h-23.7V172.4z}
 svg{M88.7,56.8c0,5.5-4.5,10.1-10.1,10.1c-5.6,0-10.1-4.6-10.1-10.1c0-5.6,4.5-10.1,10.1-10.1C84.2,46.7,88.7,51.3,88.7,56.8z};
}}
\newcommand\orcidicon[1]{\href{https://orcid.org/#1}{\mbox{\scalerel*{
\begin{tikzpicture}[yscale=-1,transform shape]
\pic{orcidlogo};
\end{tikzpicture}
}{|}}}}

\title[Environmental variation of the low-mass IMF]{Environmental variation of the low-mass IMF}

\author[T S Tanvir et al.]{
Tabassum S. Tanvir,$^{1}$\thanks{E-mail:\href{mailto:tabassum.tanvir@anu.edu.au}{tabassum.tanvir@anu.edu.au}}
Mark R.~Krumholz,$^{\orcidicon{0000-0003-3893-854X}\,1,2}$\thanks{E-mail: \href{mailto:mark.krumholz@anu.edu.au}{mark.krumholz@anu.edu.au}}
Christoph Federrath$^{\orcidicon{0000-0002-0706-2306}\,1,2}$\thanks{E-mail: \href{mailto:christoph.federrath@anu.edu.au}{christoph.federrath@anu.edu.au}}
\\
$^{1}$Research School of Astronomy and Astrophysics, Australian National University, Canberra, ACT~2611, Australia\\
$^{2}$Australian Research Council Centre of Excellence in All Sky Astrophysics (ASTRO3D), Canberra, ACT~2611, Australia
}

\date{Accepted XXX. Received YYY; in original form ZZZ}

\pubyear{2015}

\begin{document}
\label{firstpage}
\pagerange{\pageref{firstpage}--\pageref{lastpage}}
\maketitle

\begin{abstract}
We use a series of magnetohydrodynamic simulations including both radiative and protostellar outflow feedback to study environmental variation of the initial mass function. The simulations represent a carefully-controlled experiment whereby we keep all dimensionless parameters of the flow constant except for those related to feedback. We show that radiation feedback suppresses the formation of lower mass objects more effectively as the surface density increases, but this only partially compensates for the decreasing Jeans mass in denser environments. Similarly, we find that protostellar outflows are more effective at suppressing the formation of massive stars in higher surface density environments. The combined effect of these two trends is towards an IMF with a lower characteristic mass and a narrower overall mass range in high surface density environments. We discuss the implications for these findings for the interpretation of observational evidence of IMF variation in early type galaxies.
\end{abstract}


%%%%%%%%%%%%%%%%%%%%%%%%%%%%%%%%%%%%%%%%%%%%%%%%%%

%%%%%%%%%%%%%%%%% BODY OF PAPER %%%%%%%%%%%%%%%%%%

\section{Introduction}
\label{sec:intro}

The initial mass function (IMF) describes the mass distribution of stars at formation. It plays a critical role in a wide range of fields, from galaxy formation and evolution to nuclear astrophysics to planetary science. For this reason, measuring the IMF and understanding its origin has been one of the central issues in the study of the star formation process. \citet{1955ApJ...121..161S} first attempted to measure the IMF by using Solar neighbourhood stars. He adopted a single slope power-law distribution of the form $dN \propto M^{-\alpha} dM$, and found a best fitting slope $\alpha = 2.35$. Six decades later, this distribution is still considered the standard for stars > 1 M$_{\odot}$. However, as this function approaches zero, it diverges, indicating that there must be a turnover in the IMF at lower masses.
\citet{1979ApJS...41..513M} introduced the idea of a log-normal distribution between 0.1 and $\rm \simeq 30$  M$_{\odot}$, motivated by a clear flattening in the observed Solar neighbourhood mass distribution below 1 M$_{\odot}$. Today, the two most widely used forms of the IMF are the broken power-law distribution introduced by \citet{2001MNRAS.322..231K}, and the combination of log-normal (at low mass) and power law (at high mass) distribution by \citet{2005ASSL..327...41C} -- see \citet{2014prpl.conf...53O} for a recent review. Both of these forms feature a broad plateau at masses from $\approx 0.1-1\rm M_\odot$, and then a power-law decline at higher masses.

From a theoretical point of view the question is what physical processes are responsible for setting the shape of the IMF. We know from observations that the IMF is close to universal in the Milky Way and nearby local galaxies \citep[e.g.,][]{Lee20b}, which suggests a simple and universal physical mechanism. Turbulence is a natural choice, and there have been many theoretical models of the IMF based on the properties of turbulence. For example, \citet{1997MNRAS.288..145P} and  \citet{2002ApJ...576..870P} propose a model in which interacting shocks produce dense filaments and sheets where star formation occurs; the mass function is determined by the distribution of post-shock properties, which is a function of turbulent power spectrum. \citet{2008ApJ...684..395H, Hennebelle09a} propose a model based on the  \citet{1974ApJ...187..425P} formalism, in which the IMF is treated as a barrier-crossing problem: turbulence generates a spectrum of density fluctuations at different scales, and when one of these fluctuations crosses the barrier to becoming self-gravitating, it collapses to form a star. \citep{2012MNRAS.423.2037H, Hopkins13a} develops a model based on excursion set theory, for which the basic physical setup -- fluctuations induced by turbulence leading to barrier crossing -- is quite similar to that of \citeauthor{2008ApJ...684..395H}, but where some of the detailed assumptions, for example about the scale-dependence of the density statistics, are different.

A common feature of these models is that they are based on isothermal gas under the influence of turbulence. While such a theory can very naturally explain the power-law tail of the IMF, and there is good numerical evidence that turbulence does indeed determine this feature \citep{Nam21a}, isothermal turbulence is scale-free over the inertial range \citep{Kolmogorov1941c,Burgers1948}, with characteristic lengths only at the driving, dissipation, and sonic scales \citep{FederrathEtAl2021}. It therefore cannot explain why the IMF turns over at a particular mass that is independent of the total or Jeans mass of the parent star-forming cloud \citep{2014PhR...539...49K}. Again, numerical evidence supports this conclusion. \citet{2016MNRAS.458..673G, 2018MNRAS.480..182G} show that fragmentation of non-magnetised isothermal gas produces a pure power-law mass distribution, while \citet{Guszejnov20a} show that magnetised isothermal turbulence leads to a mass function with a peak that is simply proportional to the initial sonic mass, and therefore not plausibly related to the nearly-universal peak observed in the local IMF. These findings strongly suggest that any viable theory for the non-power-law low-mass IMF requires a deviation from the assumption of isothermal gas (however, see \citealt{Haugbolle18a} for a contrary view).

The question then becomes what physical process induces the deviation from isothermality. Deviations can occur due to gas becoming optically-thick to its own cooling radiation at a density of $n \sim 10^{10}$ cm$^{-3}$ \citep{Low76a, Rees76a, 1998ApJ...495..346M} or the onset of dust-gas coupling at a density $n \sim 10^{5}$ cm$^{-3}$ \citep{2000ApJ...538..115S,2005MNRAS.359..211L,JappsenEtAl2005,2008ApJ...681..365E}. The former leads to a mass scale that is plausibly associated with the smallest possible brown dwarfs, $\sim 10^{-3}$ M$_\odot$, but not plausibly associated with the observed IMF peak. The latter is more promising in terms of the mass scale, but \cite{2016MNRAS.458..673G} show that if one modifies hydrodynamic simulations simply by introducing an EOS stiffening at some density, the resulting IMF is extremely sensitive to its initial condition, which renders these models unable to provide the observed universal mass scale.

For this reason many authors have considered stellar radiation feedback as a potential mechanism to set a characteristic mass scale. In this picture, low mass young stars radiate due to accretion, raising the surrounding gas temperature and therefore the Jeans mass, strongly suppressing fragmentation \citep{2006ApJ...641L..45K,Krumholz11e, 2009ApJ...703..131O, Bate09a, Bate12a, Krumholz12b, Krumholz16c, FederrathKrumholzHopkins2017, 2018MNRAS.476..771C, 2020MNRAS.496.5201M, 2021MNRAS.507.2448M}. This effect suppresses the formation of low-mass objects by ensuring that the mass around them is unable to fragment further, and is instead available to accrete. Additional important physical processes that control the peak of the IMF include a combination of magnetic fields and protostellar jet feedback \citep[e.g.,][]{2004ApJ...601..930S, FederrathEtAl2014, 2021MNRAS.507.2448M}, and possibly tidal effects of the first Larson cores \citep[e.g.,][]{2018A&A...611A..89L, 2019ApJ...883..140H}.

However, there have been limited efforts thus far to determine the implications of these models for variation in the location of the IMF peak with star-forming environment. \citet{2008Natur.451.1082K} argue from analytic models, and \citet{2010ApJ...713.1120K} and \citet{Myers11a} confirm with simulations, that higher surface densities make it easier to form massive stars on the tail of the IMF, because higher optical depths trap radiation more effectively, suppressing fragmentation and allowing massive cores to undergo a processes closer to monolithic collapse. However, neither of these studies address the peak of the IMF, as opposed to the high-mass tail. \citet{2022MNRAS.509.1959S} do study the IMF peak, and predict a lower characteristic mass in metal-rich regions at high gas pressure / surface density, but these analytic model have yet to be checked by simulations, and in any event they predict only the approximate location of the IMF peak, not the full functional form of the IMF.

The problem of environmental variation is becoming urgent, however, because tentative evidence has started to emerge that, in the most extreme star-forming environments, the location of the IMF peak does change slightly. The strongest evidence for this shift has emerged from massive, early-type galaxies -- see \citet{Smith20a} for a recent review. In these systems, spectroscopic \citep[e.g.,][]{2010Natur.468..940V, Spiniello12a, La-Barbera13a, Conroy17a}, dynamical \citep[e.g.,][]{2012Natur.484..485C, Newman17a, Oldham18a}, and gravitational lensing \citep{Treu10a, Spiniello15a} all point to an IMF with a lower peak than is found in the Milky Way, though there remain some inconsistencies in the measurements as to exactly where and in which galaxies this shift in the IMF occurs \citep{Smith14b}. There are also claims of a shift of the peak to higher masses in low-mass galaxies \citep{Geha13a, Gennaro18a}, though in these systems the small numbers of observed stars has led to considerable debate regarding the statistical significance of the result \citep{El-Badry17a}. Regardless, the fact that evidence is emerging for variations in the location of the IMF peak represents both a challenge for theory and an opportunity, since reproducing and explaining these observations offers a strong test of models.

This paper aims to examine the effect of environmental variation on the initial mass function (IMF), focusing on the interaction of the environment with the two feedback effects that are thought to be most important near the IMF peak: radiation feedback and protostellar outflows. In order to isolate environmental effects, we identify the key dimensionless parameters that govern a star-forming system, and vary them systematically in order to perform a clean experiment that isolates the interaction of feedback and environmental effects from all other physical processes. We do so in the context of simulations that form a full star cluster with a measurable IMF, rather than just a single, massive core as in the earlier experiments of \citet{2010ApJ...713.1120K}. In \autoref{sec:Num}, we describe the numerical method and initial conditions of our simulations. In \autoref{sec:Results} we examine the results of the simulations. We discuss the implications of our findings for the IMF in general in \autoref{sec:Discussion} and we summarise in \autoref{Conclusion}.

%The aim is to explore the role of column-density in cloud fragmentation and stellar mass function while taking into account the effects of radiative feedback, protostellar outflows, and magnetic fields. The following subsection describes the initial conditions and simulation setup of the runs.

\section{Numerical Methods and Initial Conditions}
\label{sec:Num}
\subsection{Numerical Methods}

The numerical methods we use are identical to those of \citet{2014MNRAS.439.3420M} and \citet{2018MNRAS.476..771C}, and we refer readers there for a more detailed description; here we merely summarise the key points. We use the {\sc orion2} adaptive mesh refinement (AMR) code \citep{Li21b} to carry out our simulations. The code uses the scheme of \cite{2012ApJ...745..139L} to solve the equations of ideal magnetohydrodynamics coupled to self-gravity \citep{1998ApJ...495..821T,1999ASSL..240..131K} and radiation transfer \citep{2007ApJ...656..959K} in the two-temperature, mixed-frame, grey, flux-limited diffusion approximation. We use the tabulated density- and temperature-dependent Rosseland- and Planck-mean opacities provided by \citet{Semenov03a}.

The code uses sink particles \citep{2004ApJ...611..399K} to replace regions where protostars are forming and that are collapsing beyond our ability to resolve. Each of these particles runs an instance of the one-zone protostellar evolution model described by \citet{2009ApJ...703..131O}, which prescribes the instantaneous properties (radius, luminosity, polytropic index, etc.) of the star that particle represents. The luminosity becomes a source term in our radiative transfer equations, and we also include feedback due to protostellar outflows via momentum sources around each sink particle. We use the outflow model described in \cite{2011ApJ...740..107C}: when mass is accreted onto a sink particle, a fraction $f_w$ of it is ejected back into the simulation in the form of an outflow, presumably from an unresolved inner disc. The outflow material is launched with a speed $v_w$. In our simulation, we use the same wind model parameters described in \cite{2012ApJ...747...22H} and \cite{2018MNRAS.476..771C}: $f_{w} =0.3$ and $v_w = \min(v_{\rm kep}, 60\;\mathrm{km/s})$, where $v_{\rm kep}$ is the Kepler speed at the surface of the protostar. These parameters are motivated by observations of the momentum budget of protostellar outflows \citep{2000prpl.conf..867R}. %(CITE: Richer et al., 2000, Protostars and Planets IV, pg. 867).

\subsection{Initial Conditions}
All our simulations take place in a periodic domain filled with Solar metallicity molecular gas with a mean molecular weight of $2.33m_{p}$ (corresponding to a gas of 90\% H and 10\% He by number) and an initial temperature of 10 K. This combination of molecular weight and temperature corresponds to a sound speed of 0.19 km/s. 
%The average density is $\rm 6.96 \times 10^{-19} gcm^{-3}$ and the length of the domain is 0.46 pc. Therefore, the total mass stands at $\rm \approx 1000 M_{\odot}$. Based on the mean density the gravitational freefall time is calculated based on equation \ref{eqnff}. 
%\begin{equation}
% \rm t_{ff}=\sqrt{\frac{3\pi}{32G\rho}}
% \label{eqnff}
%\end{equation}
%Based on equation \ref{eqnff} for a mean density $\rm 6.96 \times 10^{-19} gcm^{-3}$ is 80 kyr. 
Based on the findings of \citet{2008Natur.451.1082K}, \citet{2010ApJ...713.1120K}, and \citet{Myers11a}, gas surface density is a key environmental parameter affecting the IMF, and we therefore wish to explore the effect of varying it. This in turn requires that we vary the mass and size of the box. However, since our goal is to conduct a clean experiment where the \textit{only} possible reason for differences in the outcome is feedback, we must vary these dimensional parameters in such a way as to ensure that the simulations remain re-scaled versions of one another, i.e., so that all the dimensionless numbers describing the system \textit{in the absence of feedback} (Mach number, virial parameter, etc.) remain unchanged. For a box of material with mean density $\rho$, length $L$, and mean magnetic field strength $B$, the re-scaling that accomplishes this is \citep{2014PhR...539...49K}
\begin{eqnarray}
\rho' & = & f\rho \\
L' & = & f^{-1/2} L \\
B' & = &f^{1/2} B.
\end{eqnarray}
This transformation leaves the Mach number, Alfv\'en Mach number, virial parameter, number of Jeans masses, and all other dimensionless parameters of the box unchanged for an arbitrary positive number $f$. However, this rescaling does change the surface density by a factor $(\rho'/\rho)(L'/L) = f^{1/2}$; it therefore changes the optical depth through the domain, and thus the response of the gas to feedback. Rescaling also changes the mass contained in the box, by a factor $(\rho'/\rho)(L'/L)^{3}= f^{-1/2}$, and the free-fall time $t_{\rm ff} = \sqrt{3\pi/32 G \rho}$, by a factor $f^{-1/2}$.

For our study, we consider three different boxes: a ``medium'' density box (run M hereafter) containing a mass $M_{\rm box} = 1000$ M$_\odot$ with surface density $\Sigma = 1$ g cm$^{-2}$ (corresponding to a mean density $\rho_0=7.0\times 10^{-19}$ g cm$^{-3}$, box length $L = 0.46$ pc), and ``low'' and ``high'' density boxes (L and H) hereafter, which are rescaled versions of run M with $f=1/4$ and $f=4$, respectively. \autoref{table:simparam} summarises the three simulation types. The medium density case is roughly intended to correspond to regions of vigorous massive star cluster formation found in the Galaxy; the size and surface density are comparable to those observed for cold ATLASGAL clumps \citep{Urquhart18a}. The low-density case is intended to more closely represent a nearby low-mass star-forming region such as Perseus of Ophiuchus, while the high-density case is intended to mimic the formation environments characteristic of extra-galactic super star clusters such as R136 or the protoclusters of NGC 253 \citep{Leroy18a}.

\begin{table*}
\centering
\begin{tabular}{ccccccccccccccc}
\hline \hline
Name & $M_{\rm box}$ (M$_{\odot}$) & $L$ (pc) & $\rho_0$ (g cm$^{-3}$) & $\sigma_v$ (km s$^{-1}$) &
$B_0$ (mG) &
$t_{\rm ff}$ (kyr) & N & $\mathcal{L}_{\rm max}$ & $\Delta x$ (AU) & $\rm t_{\rm cross}$ (Myr) & $\Sigma$ (g cm$^{-2}$) \\
\hline \hline
L & 2000 & 0.92 & $\rm 1.74\times 10^{-19}$& 2.4 & 0.36 & 160& 512 & 2 & 46& 0.4 & 0.5\\
M & 1000 & 0.46 & $\rm 6.96\times 10^{-19}$ &2.4 & 0.73 & 80& 512 & 2 & 23& 0.2 & 1 \\
H & 500 & 0.23 & $\rm 2.784\times 10^{-18}$ &2.4 & 1.45 &40 & 512 & 2 & 11 & 0.1 & 2 \\
\hline \hline
\end{tabular}
\caption{Simulation parameters, from left to right: run name, mass in the computational box, size of the computational box, mean density in the computational box, 3D velocity dispersion, initial magnetic field strength, mean free-fall time, number of cells per linear dimension on the base grid, maximum level of refinement, smallest cell size, turbulent crossing time, and surface density.}
\label{table:simparam}
\end{table*}

All our simulations use gravity and magneto-hydrodynamics with periodic boundary conditions. For the radiation field, we use Marshak boundary conditions, whereby radiation can freely leave the domain, but the domain is also bathed with an external radiation field that corresponds to a 10~K isotropic blackbody. 

The simulation itself is divided into two phases. In the first phase, which begins with uniform density and magnetic field, we turn off gravity and radiation, 
%turn on turbulence driving, and 
set $\rm \gamma = 1.0001$ so that the gas is close to isothermal, and drive turbulence in the box using the method of \cite{1999ApJ...524..169M}. The driving field is purely solenoidal \citep{FederrathKlessenSchmidt2008}, with constant power per wave number $P(k)$ over the range $1 \leq k L/2\pi \leq 2$. At every time step we apply a force with a spatial distribution matching the driving spectrum, and with an amplitude chosen to maintain a roughly constant mass-weighted RMS thermal Mach number of $\mathcal{M} = \sigma_v/c_{s} = 12.6$ where $\sigma_v$ is the three-dimensional (3D) velocity dispersion. The corresponding virial parameter is \citep[e.g.,][]{FederrathKlessen2012}
\begin{equation}
\alpha = \frac{5\sigma_{v}^{2}R}{GM_{\rm box}} = 1,
\label{eqnvir}
\end{equation}
where we have taken the characteristic radius $R=L/2$. The corresponding turbulent crossing time is
\begin{equation}
t_{\rm cross} = \frac{L}{\sigma_v} = (0.4, 0.2, 0.1) \mbox{ Myr}
\end{equation}
for L, M, and H runs, respectively. 

In addition to $\mathcal{M}$ and $\alpha$, the final dimensionless number that characterises our simulation is the plasma $\beta$. We fix this parameter by initialising the uniform magnetic field at the start of the first phase of the simulation to the value $B_{0}$ required to have \citep[e.g.,][]{FederrathKlessen2012}
\begin{equation}
\beta = \frac{8\pi \rho c_{s}^{2}}{B_{0}^{2}}=2\left(\frac{\mathcal{M_{\rm A}}}{\mathcal{M}}\right )^{2} = 0.012
\end{equation}
The corresponding Alfv\'en Mach number $\mathcal{M}_{\rm A}$ = 1.
As noted above, by fixing $\mathcal{M}$, $\alpha$, $\mathcal{M}_{\rm A}$, and $\beta$, we also fix all other dimensionless numbers in the absence of radiation and feedback.

%[NAME OF WHICH MAGNETIC PARAMETER YOU ARE FIXING -- $\mathcal{M}_{\rm A}$, $\beta$, $\mu_\Phi$, etc.] We fix this parameter by initialising the uniform magnetic field at the start of the first phase of the simulation to the value $B_0$ required to have [EQUATION OF THE FORM \textbf{PARAMETER} = \textbf{DEFINITION OF PARAMETER IN TERMS OF $B_0$} = \textbf{NUMERICAL VALUE}]

%We define the magnetic field strength in terms of plasma $\beta$ parameter
%\begin{equation}
%\beta = \frac{8\pi \rho c_{s}^{2}}{B_{rms}^{2}}
%\end{equation}
%where $B_{rms}$ is the root mean square volume weighted magnetic field strength. The Alfv\'en mach number is defined as 
%\begin {equation}
%\mathcal{M}_{A} = \frac{v_{\rm rms}}{v_{\rm A}}
%\end {equation}
%where $v_{\rm A}$ is the Alfv\'en speed,$v_{\rm A} \equiv B_{\rm rms}^{2}/4\pi \rho$. 

We evolve the initially uniform gas with driving for two crossing times, and then treat the cloud state at the end of the two crossing times as the initial conditions for the turbulent cluster simulation. At this point we turn on gravity, sink particles, and radiation and turn off turbulence driving. We also set $\rm \gamma = 5/3$ instead of 1.0001, which allows the temperature to vary with the outcome of the radiative transfer calculations. We run these simulations until they reach 5$\rm \%$ star formation efficiency, i.e., until 5\% of the initial gas has been accreted by star particles. 
%The resolution of the protostellar cluster simulation is discussed in the following section.

\subsection{Resolution, refinement, and sink particles}

We initialise the AMR hierarchy on a $512^{3}$ base grid which we denote as $\rm \mathcal{L} = 0$. During the driving phase, we disable AMR, so no higher levels exist. We use a 512$^3$ grid during this phase because this resolution is sufficient to capture a well-resolved inertial cascade down to the scale of $L/30$ \citep{2010A&A...512A..81F,2011ApJ...731...62F,2012ApJ...745..139L}. This scale is one order of magnitude less than the initial Jeans length in our simulations, so during the driving phase this choice ensures that we have well-resolved turbulence down to scales well below the scale at which self-gravity is expected to become important.

Once we turn on gravity we allow the grid to adaptively refine to a maximum level $\rm \mathcal{L}_{max} = 2$. We use two different criteria to decide where to add resolution. First, we increase the resolution wherever fewer than eight zones resolve the local Jeans length, $\rm J > 1/8$ \citep{1997ApJ...489L.179T}. This criterion triggers refinement to the finest level at a density of 1/4 of that required to trigger the insertion of a sink particle at the finest level, thereby ensuring that strongly collapsing regions get refined to the finest level. The second condition is that we refine any cell that contains a poorly resolved gradient of the radiation energy density, $|\nabla E_{r}|\Delta x/E_{r} > 0.125$, where $\Delta x$ is the grid spacing. This ensures that we resolve the radiation field and temperature structure. Our choice $\mathcal{L}_{\rm max} = 2$ corresponds to a maximum resolution $\Delta x = 23$ AU for the medium run, and factors of two smaller (larger) for the high (low) surface density runs. 
%For a background temperature of 10K, the density threshold at which sink particle formation is triggered for an effective resolution of 23 AU is $8.8 \times 10^{-15}$ g cm$^{-3}$.

In our simulations, sink particles form in any zone on the finest AMR level where the gas becomes dense enough to reach a local Jeans number
\begin{equation}
    J = \sqrt{\frac{G\rho \Delta x^{2}}{\pi c_{s}^{2}}} > \frac{1}{4}.
\end{equation}
At a temperature of 10~K this corresponds to a density $\rho_{\rm sink} = 8.8 \times 10^{-15}$ g cm$^{-3}$ for run M, and a factor of 4 lower (higher) for run L (H). However, in practice the gas near the sink creation threshold is often significantly warmer than 10~K once the simulation is underway, and thus the actual maximum resolvable density is substantially higher.
Once formed, these sink particles evolve and interact with the gas via gravity and accretion following the procedure described in \cite{2004ApJ...611..399K}. This has been updated to include the effects of the magnetic field on the gas accretion rate on sink particles \citep{2014ApJ...783...50L}.

There is a limitation in the method we use for radiation transfer: we assume that the gas and dust temperatures are equal. Since the gas and dust temperatures become identical at densities above $\rm 10^{4}-10^{5} cm^{-3}$ \citep[e.g.,][]{2001ApJ...557..736G}, this is a good assumption for most of the gas in the simulation box. However, in the low density, non-self-gravitating regions, the density may be too low to allow efficient gas-grain coupling, allowing gas to be either hotter or cooler than the dust, depending on the environment. However, this is unlikely to be important for the purpose of determining the IMF, since in the regions where gas is collapsing and fragmenting, the density is high enough that grains and gas are well-coupled. Simulations that include an explicit treatment of imperfect grain-gas coupling \citep{Bate15a} find that its effects on fragmentation are minimal \citep{Bate19a}.

\subsection{Parameter study}

We carry out two different realisations for each simulation type (L, M, and H). For a given simulation type, the two realisations are identical in their mean properties and differ only in their randomly-generated turbulent driving fields. We refer to simulations using the first random realisation as L1, M1, and H1, and those using the second random realisation as L2, M2, and H2. Runs L1, M1, and H1 all use the same driving pattern, as do runs L2, M2, and H2. We also carry out one additional simulation, ISO1, which uses the same driving pattern and initial conditions as M1, but where we have disabled both outflow feedback and radiative transfer, and we continue to use an isothermal equation of state ($\gamma = 1.0001$) throughout the collapse phase.

We emphasise that since L1, M1, H1, and ISO1 all use the same driving field (as do L2, M2, and H2), have identical dimensionless parameters, use the same initial and maximum resolution, and use the same Jeans refinement criterion (in the sense that the factor by which a given cell must be overdense relative to the box mean in order to be flagged for refinement is the same in all cases), the initial conditions and simulation setup are \textit{identical to machine precision}. The only physics elements of L1, M1, H1, and ISO1 that are not identical are radiation and outflow feedback;\footnote{Exact identity to machine precision is also broken slightly by differences in numerical truncation error caused by the fact that the time steps in the different runs are not completely identical. However, this is not a significant physical effect.} these break the symmetry between the runs, and introduce characteristic length and mass scales that would not exist in their absence. Any differences in outcome between the various runs must therefore be due to one of these factors.

\section{Results}
\label{sec:Results}
Here we present the results of our simulations. 
%The result section is divided into 4 subsection. in section 3.1 we discuss the global evolution of our gas clumps, the star formation activity. 
We first give an overview of the global evolution of the simulations in \autoref{ssec:overview}.
%The following section 
In \autoref{ssec:mass_dist}
we discuss the sink particle mass distribution, and identify differences between the various runs. In \autoref{ssec:rad_feedback} and \autoref{ssec:outflow_feedback} we discuss the roles of radiation feedback and protostellar outflows, respectively, in shaping the mass distribution.

\subsection{Overview of simulations}
\label{ssec:overview}
\begin{figure*}
	\includegraphics[width=\textwidth]{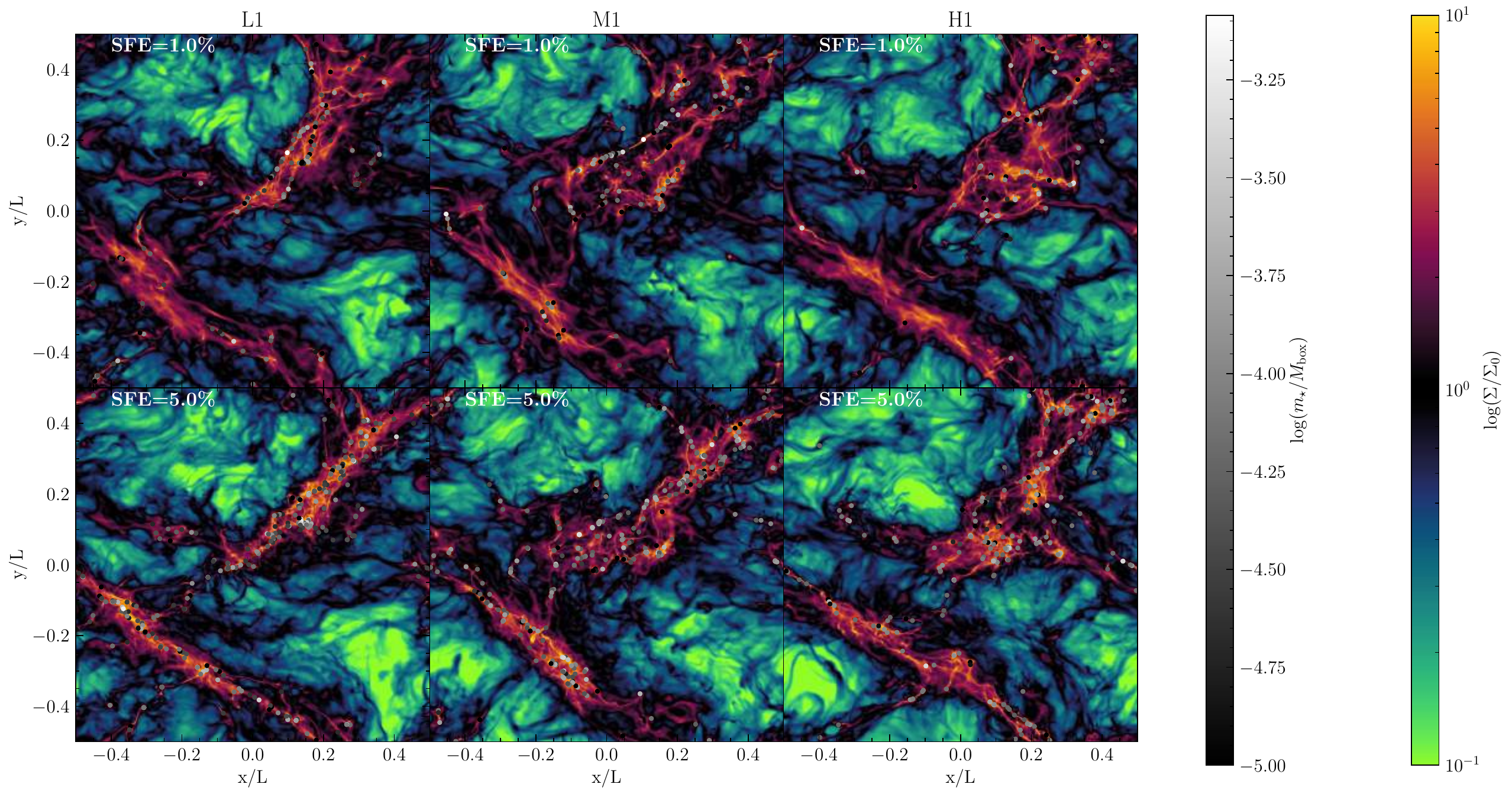}
    \caption{Column densities in simulations L1, M1, and H1 (left to right). The top row shows the simulation state at $1\%$ star formation efficiency, the bottom row at $5\%$ star formation efficiency. The colour scale goes from $\log(\Sigma/\Sigma_{0}) = -1$ to $1$, where $\Sigma_0 = \rho_{0} L$ and $\rho_0$ is the mean density in the simulation domain. Circles show star particles, and are colour-coded by mass $m_\star$ from $\log (m_{\star}/M_{\rm box}) = -5 $ to $-3$, where $M_{\rm box}$ is the total mass of the simulation box.}
    \label{fig:proj}
\end{figure*}

\begin{figure*}
	\includegraphics[width=\textwidth]{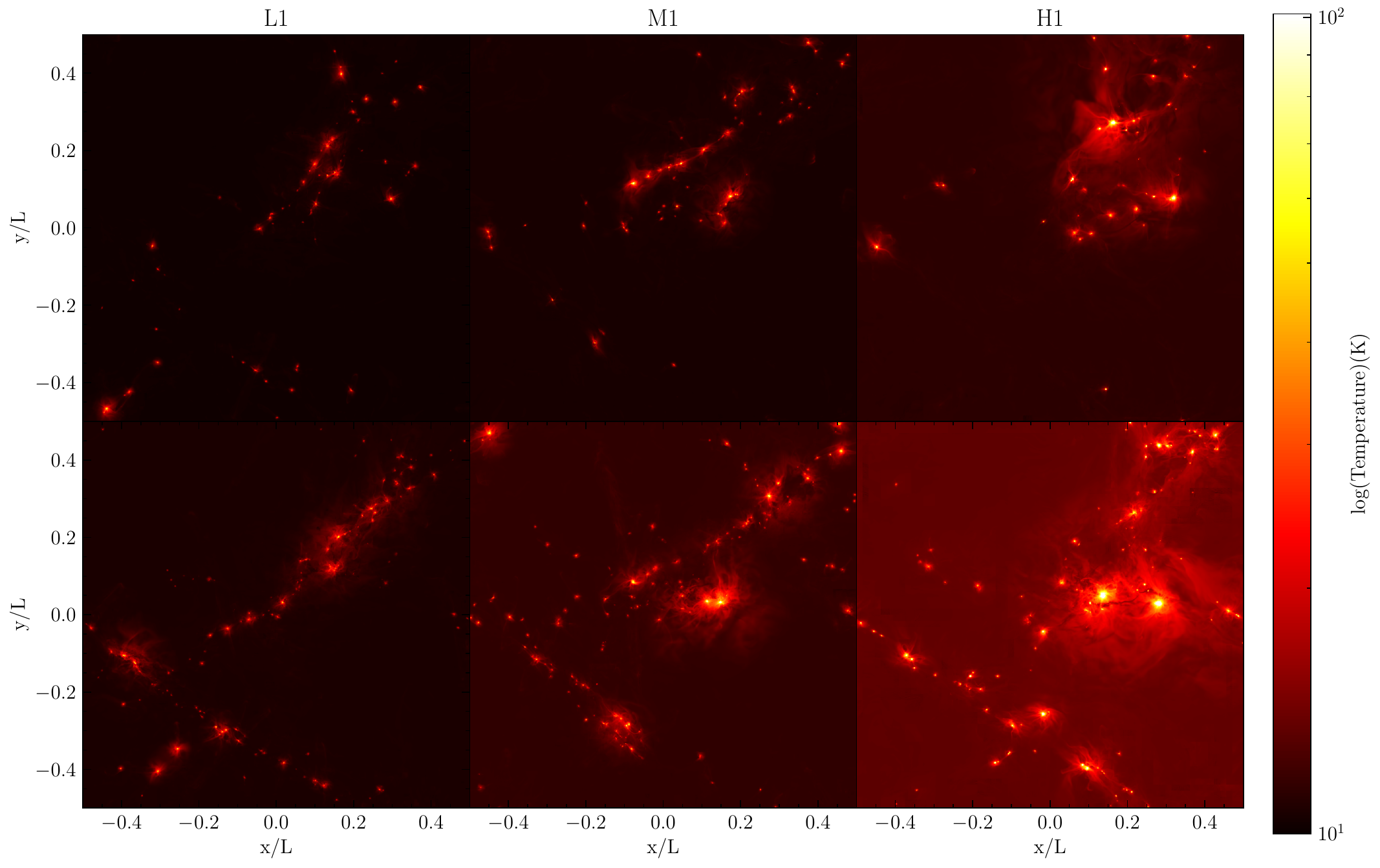}
    \caption{Same as \autoref{fig:proj}, but showing density-weighted projected temperature rather than column density.}
    \label{fig:projtemp}
\end{figure*}

In \autoref{fig:proj} and \autoref{fig:projtemp} we show the column density and density-weighted mean temperature of runs L1, M1, and H1, one of our two turbulent realisations; while the exact morphology is of course different for realisation 2, the qualitative discussion we present here applies equally well to that case. Since star formation occurs at slightly different rates in each of these simulations, we show results at matching star formation efficiencies rather than matching times, though the difference is relatively small (provided time is measured in units of the simulation free-fall time). In all three runs the turbulence has created dense filamentary structures, which over time further fragment into isolated gas cores that collapse to form stars. As can be seen from \autoref{fig:proj}, star formation activity is mostly confined in the filamentary regions. The overall morphologies of the L1, M1, and H1 runs are very similar, which is not surprising since, absent the effects of feedback, the runs would be identical. However, some small differences are clearly visible: as the column density of the runs increases the filamentary shapes become more disrupted; filaments are narrower and straighter in run L1, and broader and more dispersed in run H1.

To better understand the collapse morphology, fragmentation and star formation we look at the temperature structure of the gas shown in \autoref{fig:projtemp}. This shows much larger differences between the runs than \autoref{fig:proj}. Compared to run L1, run H1 is warmer because it has a higher surface density, which leads to a higher optical depth that traps the radiation produced more effectively. The difference between the runs increases with star formation efficiency as the influence of radiation feedback on the temperature grows.

\begin{figure}
	\includegraphics[width=1.0\columnwidth]{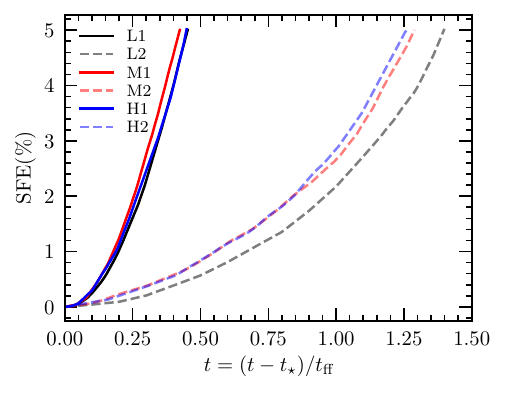}
    \caption{Star formation efficiency as a function of time  since the formation of the first star $t_{\star}$, measured in units of the free-fall time $t_{\rm ff}$. The solid lines are realisation 1 and the dashed lines are for realisation 2.}
    \label{fig:sfetime}
\end{figure}

\begin{figure}
	\includegraphics[width=1.0\columnwidth]{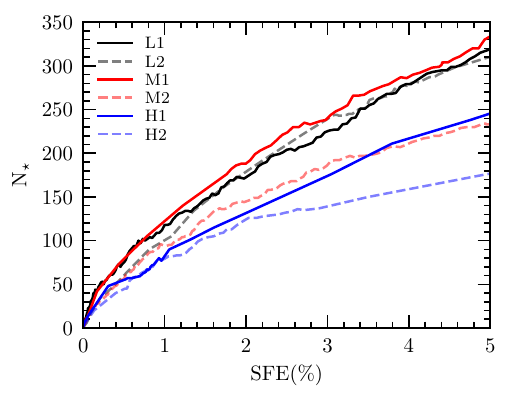}
    \caption{Number of stars formed as a function of star formation efficiency. The solid lines are realisation 1 and the dashed lines are realisation 2.}
    \label{fig:nstarssfe}
\end{figure}

%\subsection{Star formation activity}
%In this section 
We show the time evolution of the star formation efficiency (SFE) in \autoref{fig:sfetime}, and the total number of stars present as a function of the SFE in \autoref{fig:nstarssfe}. We define the SFE as the fraction of initial box mass converted to stars, i.e.,
\begin{equation}
 \mathrm{SFE} = \frac{m_{\star}}{M_{\rm box}} = \frac{m_{\star}}{M_{\rm gas}+m_{\star}}
\end{equation}
where $m_{\star}$ is the total stellar mass and $M_{\rm box}$ is the total of the gas and stellar mass in the simulation (which is always equal to the initial mass in the box). From the onset of star formation (which is itself $\sim 0.5$ free-fall times after we turn on self-gravity), it takes approximately $0.5 t_{\rm ff}$ for run L1 to reach 5\% SFE. For run L2 it takes almost 3 times as long to reach the same SFE. The difference is simply due to the different random realisations in the driving pattern, and is within the normal range of variation with driving pattern (e.g., see Figure 4 of \citealt{Nam21a}). By contrast, there is fairly little difference between runs with the same driving pattern but different surface densities. This suggests that, whatever effects feedback is having in reducing the SFE, those effects are almost independent of surface density.

If we examine \autoref{fig:nstarssfe}, which shows the number of stars versus SFE, we see a somewhat different pattern. The differences between the two realisations (solid versus dashed lines) are relatively modest, but there is a clear trend with surface density: at fixed SFE, the high surface density run has produced the fewest stars, and the low surface density run the most. The one exception to this is runs L1 and M1, where the numbers of stars are quite similar. Thus, we see that the SFE does not depend strongly on the absolute surface density (in our scaled experiments), but the number of stars at fixed SFE, and by implication the stellar mass distribution, does.
%we can see that both runs have produced nearly the same number of stars in both cases. This however is not true for run M1 and M2. Like run L2, run M2 also takes longer to convert the same amount mass into stars but run M2 produces less stars than run M1. This difference in the star formation activity in these runs are purely down to their randomly generated turbulent driving field.
%plots of star formation efficiency (SFE) versus time and star formation efficiency (SFE) versus the total number of stars 

\subsection{Stellar mass distribution}
\label{ssec:mass_dist}
%In this section we examine the evolution of the distribution of sink particle mass of all the simulation runs. 
We now examine the stellar mass distributions in the simulations more closely.
In all the figures we present in this section, we show results using two different mass scales: absolute, and relative to box mass (i.e., stellar masses are expressed as $m=m_{\star}/M_{\rm box}$). In the absence of both radiation feedback and outflow feedback the relative mass distribution would be identical for all the simulations, and the absolute mass distributions would simply scale as $f^{-1/2}$, i.e., we should find that the absolute mass distributions in the H runs have the same shape as in the corresponding M runs, but shifted to lower masses by a factor of 2; similarly, L runs should be shifted a factor of 2 higher in absolute mass than M runs.

\begin{figure}
	\includegraphics[width=1.0\columnwidth]{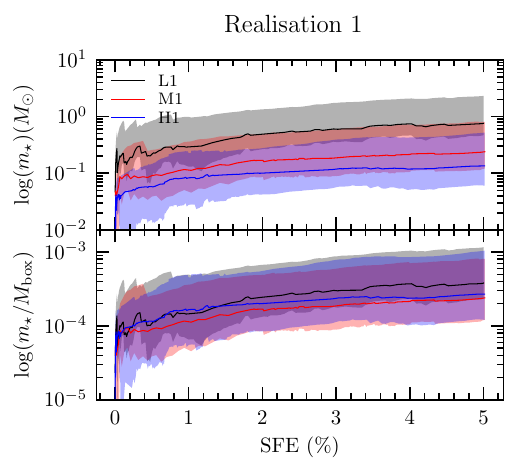}
	\includegraphics[width=1.0\columnwidth]{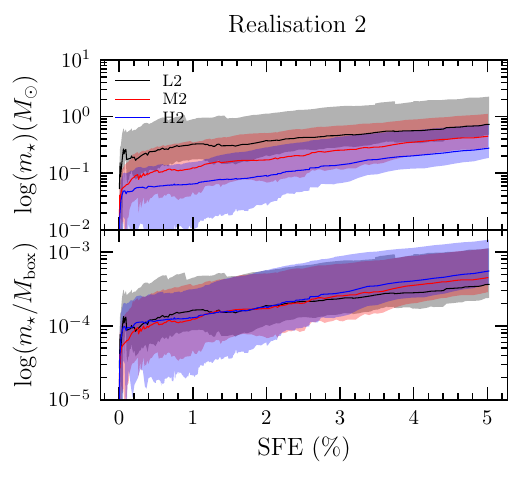}
    \caption{Solid lines show the evolution of the median of the sink particle mass distribution for realisation 1 (top two panels) and realisation 2 (bottom two panels), while the shaded regions around them show the 25th to 75th percentile range; here medians and percentiles are measured with respect to mass rather than number, i.e., the mass $m_\star$ plotted is the one for which half the stellar mass is in stars with masses $<m_\star$. In each pair of panels, the upper one shows the absolute mass expressed in M$_\odot$, while the lower one shows the mass expressed relative to the box mass $M_{\rm box}$. The black, red, blue lines and shaded regions correspond to the L, M, and H runs, respectively.}
    \label{fig:median}
\end{figure}

We begin by looking at the two sets of plots in \autoref{fig:median}, which shows the evolution of the median and 25th to 75th percentile ranges of the sink particle mass distributions for all sets of simulations. Here we measure the median and percentiles with respect to stellar mass rather than number, i.e., if $m_{25}$ is the 25th percentile mass, this means that the stars with masses $m<m_{25}$ together constitute 25\% of the stellar mass in the simulation at that time, not that they are 25\% of the stars by number.\footnote{We prefer to work with percentiles computed by mass rather than number, because the latter are quite sensitive to the numerical details of the sink particle algorithm, such that different choices for parameters such as the Jeans number used when calculating the sink particle creation density threshold can lead to different numbers of very low-mass objects (e.g., see Appendix C of \citealt{Haugbolle18a}). Percentiles computed by mass are far less sensitive to such numerical details.} From the plots we can see that the median masses are relatively converged by the time we reach 5$\%$ SFE, and are increasing only very slowly. Though most of the sink particles at 5$\%$ SFE are still accreting, this is counterbalanced by the fact that new sink particles are forming at the same time. Therefore, the population as a whole has reached nearly a steady state distribution. When measured on an absolute scale (top panels), we see that the median mass, along with the 25th to 75th percentile range, decreases as the surface density increases, as we would expect. However, the decrease is not by exactly the amount that would be expected for scale-free behaviour; when measured relative to the box mass, the stellar mass distributions are not identical but instead show clear variations.

\begin{figure}
	\includegraphics[width=1.0\columnwidth]{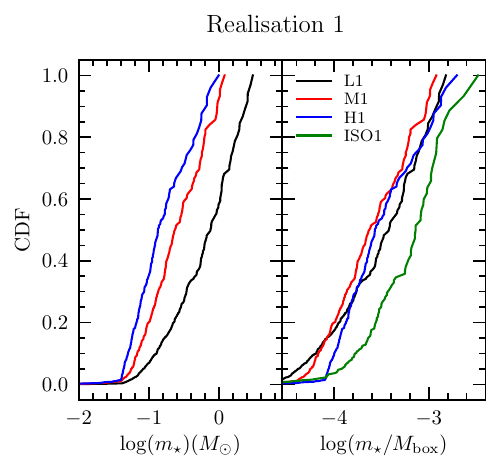}
	\includegraphics[width=1.0\columnwidth]{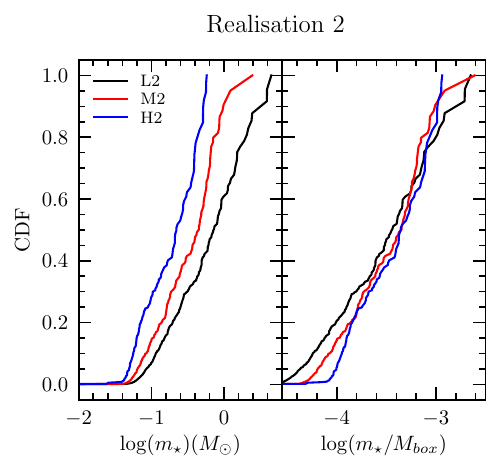}
    \caption{Cumulative distribution function (CDF) of the sink particle masses of the simulations; we show realisation 1 in the upper pair of panels, and realisation 2 in the lower pair. The black, red and blue lines refer to the low, medium and high surface density runs. The green line denotes the run with no feedback. The left sides of both pairs of panels show the CDF with respect to absolute stellar mass, while the right sides show the CDF for mass measured relative to the box mass $M_{\rm box}$. The low surface density runs have a broader mass range and higher mean mass than the higher surface density runs. The no-feedback run is a continuation of this trend with a broader mass range and higher mean mass than the lower surface density runs.}
    \label{fig:cdf_plot}
\end{figure}

To explore these differences further, in \autoref{fig:cdf_plot} we show the cumulative mass functions of the runs at 5$\%$ SFE, again on both an absolute and a relative mass scale; as with \autoref{fig:median}, we measure the CDF with respect to mass, not number. From \autoref{fig:cdf_plot} it is clearly evident that the radiation feedback and outflow feedback have broken the symmetry between these runs. However, the effect is not simply a shift in the median, but is rather a change in the shape as well. The mass range in the higher surface density cases is narrower than for lower surface density runs. Measured relative to the box mass, we see a characteristic pattern that the higher surface density runs (blue) have fewer low-mass stars (i.e., the CDF begins to rise above zero at higher mass), but the same runs \textit{also} have fewer very high mass stars (i.e., the CDF approaches unity at lower mass). Viewed on an absolute scale, the higher surface density runs have IMFs that are both lower in mean mass and narrower in range of mass. The no-feedback run appears as a continuation of this trend, i.e., the absolute mass scale is even larger than for run L1, and the mass distribution is even broader.
% We will look at role of these effects in setting the mass functions in subsequent sections.

%In figure XXX we also show the distribution of the sink particles 

%the higher surface density runs have a much lower median mass and a steeper distribution compared to the low surface density runs.

%In this section we now focus on the mass distribution in these simulations. 
\subsection{The effect of radiation feedback on the simulations}
\label{ssec:rad_feedback}

\begin{figure*}
	\includegraphics[width=2.2\columnwidth]{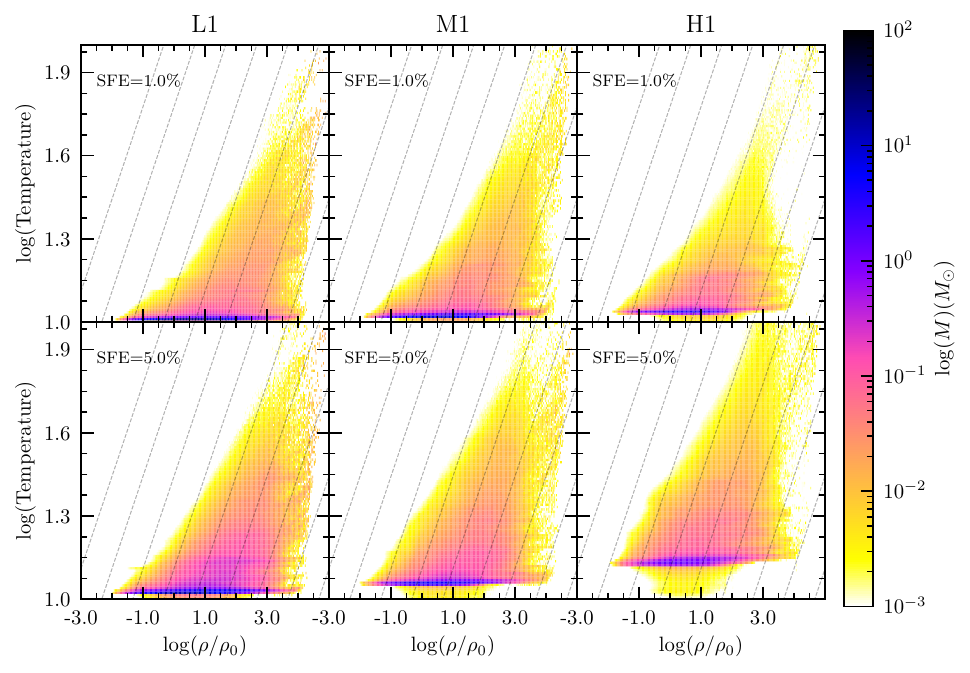}
	
    \caption{The joint distribution of normalised density $\rho/\rho_0$ and temperature $T$ for runs L1, M1, H1 (left to right). The top row shows the state of the simulations at $\rm 1 \%$ SFE and the bottom row at 5\% SFE. The colour bar shows how much mass is in each density-temperature bin. Dashed lines indicate loci of constant box-normalised Jeans mass $M_{\rm J}/M_{\rm box}$; lines are spaced logarithmically at intervals of 0.5 dex, with the left-most line corresponding to $\log M_{\rm J}/M_{\rm box} = -2$.}
    \label{fig:phase2d}
\end{figure*}

We now seek to understand the mechanisms by which feedback breaks the symmetry between our different cases and produces the differences in the IMF that we have observed. We begin in this section by studying the effects of radiation feedback, which alters the gas temperature distribution. \autoref{fig:phase2d} shows the distribution of the gas mass with both density and temperature in runs L1, M1 and H1 at SFEs of 1\% and 5\%. In this and all similar figures in this section, realisation 2 is qualitatively the same as realisation 1, so we only show plots of realisation 1 for reasons of space. From the plot it is clear that as the surface density increases the simulations become progressively warmer, therefore suppressing fragmentation. This is consistent with earlier studies of radiation feedback, both numerical \citep[e.g.,][]{2010ApJ...713.1120K, Myers11a,2020MNRAS.496.5201M,2021MNRAS.507.2448M} and analytic \citep[e.g.,][]{Chakrabarti05a, 2006ApJ...641L..45K, Krumholz11e, 2008Natur.451.1082K, 2016MNRAS.458..673G} which find that higher surface densities trap radiation more effectively, leading to systematically higher gas temperatures at fixed ratio of stellar mass to gas mass. To help visualise this effect, the dashed lines in \autoref{fig:phase2d} illustrate loci of constant box-normalised Jeans mass $M_{\rm J}/M_{\rm box}$, where 
\begin{equation}
    M_{\rm J} = \frac{\pi}{6} \left(\frac{k_B T}{G \mu}\right)^{3/2} \frac{1}{\rho^{1/2}}
\end{equation}
is the Jeans mass. It is clear that, as we move from L1 to M1 to H1, the mass distribution shifts systematically to the left relative to these lines, indicating less mass at low $M_{\rm J}/M_{\rm box}$.

\begin{figure}
	\includegraphics[width=1.0\columnwidth]{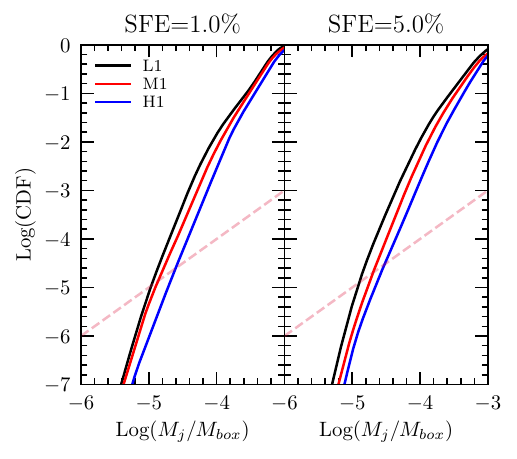}
	
    \caption{Cumulative distribution functions of $M_{\rm J}/M_{\rm box}$ in runs L1, M1, and H1 (black, red and blue lines) at SFEs of 1\% (left) and 5\% (right). The dashed one-to-one lines indicate a minimum condition for fragmentation: for any mass $M_{\rm J}/M_{\rm box}$ for which the CDF falls below the dashed line, there is less than a single mass of Jeans material in the box, and thus it is impossible to create an object of that mass via gravitational collapse.}
    \label{fig:jeansCDF}
\end{figure}

To explore this effect further, we show the 1D mass-weighted cumulative distribution function of $M_{\rm J}/M_{\rm box}$ in \autoref{fig:jeansCDF}. The way to interpret this plot is that each point on the horizontal axis corresponds to one of the dashed lines in \autoref{fig:phase2d}, with higher values of $M_{\rm J}/M_{\rm box}$ (moving right in \autoref{fig:jeansCDF}) corresponding to moving left in \autoref{fig:phase2d}. The value on the vertical axis in \autoref{fig:jeansCDF} indicates what fraction of the mass in the simulation lies to the right of the corresponding dashed line in \autoref{fig:phase2d}. We see a clear trend with surface density: in run M1, the distribution extends to lower $M_{\rm J}/M_{\rm box}$ than in L1, which in turn extends to lower values than H1.

To visualise how this trend might impact fragmentation, in the plot we also show the 1-1 relation (dashed lines). The significance of the 1-1 lines becomes clear if we recall that the CDF is also a quantity measured in units of a fraction of the total simulation mass $M_{\rm box}$. Consider for example $M_{\rm J}/M_{\rm box} = 10^{-5}$ at a SFE of 1\% (left panel). Examining the plot, we see that the CDF of run H1 is below the 1-1 line at this value of $M_{\rm J}/M_{\rm box}$; the CDF value here is $\approx 10^{-6.2}$, i.e., the amount of mass in the box for which $M_{\rm J} < 10^{-5} M_{\rm box}$ is $\approx 10^{-6.2} M_{\rm box}$. Thus, in the entire simulation box, there is only $\approx 10^{-6.2}/10^{-5} =6.3\%$ of a Jeans mass of material dense and cold enough that its Jeans mass is $<10^{-5} M_{\rm box}$ -- there is too little mass in the box to make even a single object this small via gravitational collapse. By contrast, at the same value of $M_{\rm J}/M_{\rm box}$, run L1 is above the 1-1 line, and has a CDF value of $\approx 10^{-4.8}$, so there is $1\approx 10^{-4.8}/10^{-5} = 160\%$ of a Jeans mass of material. Of course this is only a necessary condition for the creation of such a small collapsed object, not a sufficient one; the CDF is a point statistic, and tells us nothing about how this mass is arranged. It may be split up into a dozen locations spread throughout the simulation volume, such that no single contiguous region has enough mass to be gravitationally unstable and collapse. Nonetheless, it is clear that runs L1 and H1 are fundamentally different at this mass: run L1 can in principle make collapsed objects with masses as small as $10^{-5} M_{\rm box}$, while run H1 cannot. The intersection of the dashed 1-1 line with the CDF therefore represents at least a rough estimate of the minimum object mass that it is possible for gravity to produce.

Armed with this understanding, we can now make sense of the differences we observe at the low-mass ends of the IMFs produced in the different simulations. \autoref{fig:jeansCDF} shows that fragmentation to the smallest objects is suppressed more at the higher surface density runs than the lower surface density runs. Thus radiative feedback suppresses the IMFs at the low mass end, which is exactly what we observe: when measured in box-normalised units, run H has fewer very low-mass stars, and run L has more. However, we remind readers that this statement applies to masses measured in normalised units -- while radiation feedback does push the low end of the IMF upward in run H1 compared to L1, it does not do so by enough to fully compensate for the overall lower value of the mean Jeans mass in run H1 compared to L1. Thus the lowest-mass stars in run H1 are still (slightly) smaller than those in run L1 -- just by much less than the factor of four difference that would be expected given the differences in mean density between the two cases.

However, our analysis thus far does not explain what is happening at the high-mass end of the IMF. If radiation were the only actor, we would expect that at higher surface density, more mass should be accreted onto the high mass stars as well, leading the overall IMF to shift upwards (in normalised units). That is not what we observe in \autoref{fig:cdf_plot}. This suggests that outflow rather than radiative feedback is dominant at the high mass end of the IMF. We turn to this question next.
%shows the distribution of the runs 

\subsection{The effect of the outflow feedback on the simulations}
\label{ssec:outflow_feedback}

\begin{figure}
	\includegraphics[width=1.0\columnwidth]{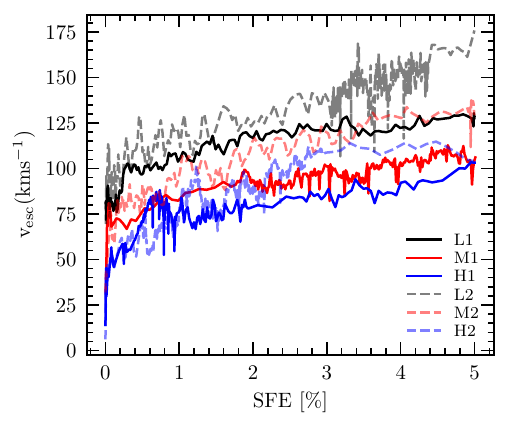}
	\caption{Accretion weighted mean surface escape speed of the simulations against SFE. The solid lines are realisation 1 and the dashed lines are realisation 2.}
    \label{fig:vesc}
\end{figure}

\begin{figure}
	\includegraphics[width=1.0\columnwidth]{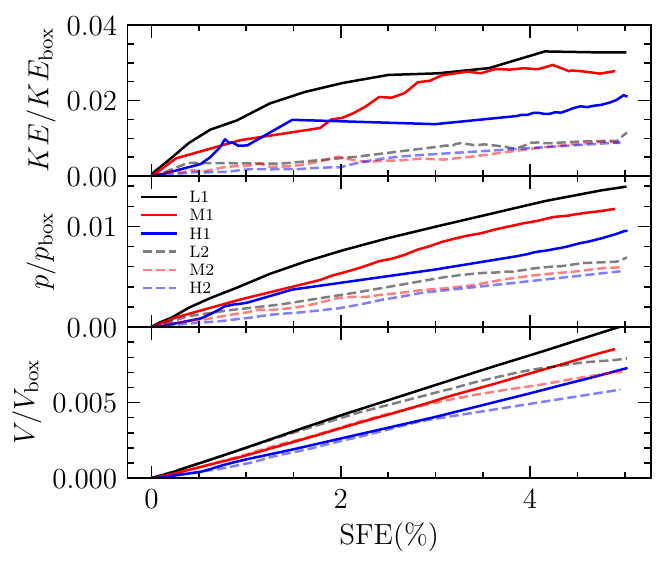}
	\caption{Outflow kinetic energy, outflow momentum and outflow volume relative to the box kinetic energy, momentum and volume, all as a function of SFE. The solid lines represent realisation 1 and the dashed lines represent realisation 2.}
    \label{fig:outflow}
\end{figure}

If radiation feedback were the only feedback mechanism operating, our analysis from the previous section suggest that we should we should see a uniform shift of the IMF towards higher (normalised) masses in the high surface density simulations. However, this is not what we observe: the higher surface density simulations produce fewer low-mass stars (i.e., the IMF is shifted upward at the low-mass end), but \textit{also} fewer -- or at least an equal number of -- high-mass stars. There is no obvious mechanism by which radiation could produce such an effect, since in the high surface density runs are uniformly warmer, and higher temperature favours more massive objects. Instead, the most likely explanation is that the decrease, or lack of increase, in the abundance of high-mass stars in the high surface density runs is that protostellar outflow feedback is breaking up the gas more effectively in these runs, making it harder for more massive objects to grow.

To understand how and why outflow feedback depends on the surface density, in \autoref{fig:vesc} we show the evolution of accretion rate-weighted mean surface escape speed of the runs as a function SFE. This quantity matters for outflows, because our outflow prescription links the outflow velocity to the stellar surface escape speed -- stars with higher escape speeds have faster outflows -- and thus the accretion rate-weighted mean surface escape speed is directly proportional to the outflow momentum flux. The correlation between surface escape speed and outflow speed implemented in our model is both seen observationally \citep[and references therein]{2000prpl.conf..867R} and expected theoretically \citep[and references therein]{Konigl00a}, since outflows are launched from close to the stellar surface; see \citet{2011ApJ...740..107C} and \citet{FederrathEtAl2014}for details.

From the plot it is clear that the lower surface density runs have systematically higher escape speed than the high surface density runs, and thus inject more outflow momentum (in normalised units). This difference occurs because in physical time units, the lower surface density runs take the longest to reach any given SFE, which gives the stars more time to contract toward the main sequence. Just as the optical depth of the domain is a quantity that is not conserved as we rescale from the L to the M to the H runs, and this breaks the symmetry with respect to radiation feedback, the number of Kelvin-Helmholtz times per free-fall time is also not conserved under rescaling, and this breaks the symmetry with respect to outflow feedback.

To see how this symmetry breaking manifests in the simulations, in \autoref{fig:outflow} we show the outflow kinetic energy, scalar momentum and volume occupied normalised to the box kinetic energy, momentum and volume; formally, we normalise the kinetic energy by $(1/2)M_{\rm box} \sigma_v^2$ (where $\sigma_v$ is the initial 3D velocity dispersion), the momentum to $M_{\rm box}\sigma_v$, and the volume to $L^3$. For the purposes of this plot, we define outflow material using \textsc{orion2}'s passive scalar capability: material that is accreted by stars and then re-inserted into the computational domain by our stellar ouflow prescription is tagged with a passive scalar that is conservatively transported thereafter, and thus in every cell at every time we know the partial density $\rho_{\rm out}$ of outflow material. The outflow kinetic energy, scalar momentum and volume in a given cell of total density $\rho$, velocity $v$, and volume $V$ are then $(1/2) \rho_{\rm out} v^2$, $\rho_{\rm out} v$, and $(\rho_{\rm out}/\rho)V$.

Examining \autoref{fig:outflow}, we see that outflows in the low surface density runs carry almost three times as much kinetic energy and twice as much scalar momentum, and occupy twice as much volume, as in the high surface density runs. Again, we remind the reader that, due to the symmetry between the L, M, and H runs, by construction the only possible sources for these differences are the differing rates of outflow momentum injection (c.f.~\autoref{fig:vesc}), or a secondary effect of the differing temperature distributions that results from the broken symmetry of stellar radiation feedback. Regardless of the source of the difference, one might expect based on \autoref{fig:outflow} that outflow feedback would be more effective in the L runs, not the H ones as we seem to require. However, the higher outflow occupation volume, coupled with the more fragmented and less filamentary density distribution we see in \autoref{fig:proj}, suggests a solution. The greater outflow injection momentum in the low surface density runs makes it easier for the outflows to punch holes in the collapsing cores in these cases. The outflows rapidly break out of the dense gas, transferring less of their momentum in the process, and more easily reaching low-density regions where they are able to expand and occupy more volume. This leads to less fragmentation of the dense gas in the low surface density runs, and therefore makes outflow feedback less effective, leading to a heavier IMF at the massive end.

%low surface densities have a higher escape speed than the high surface density runs. 

\section{Discussion}
\label{sec:Discussion}

\subsection{Implications for IMF variations}

\begin{figure}
	\includegraphics[width=1.0\columnwidth]{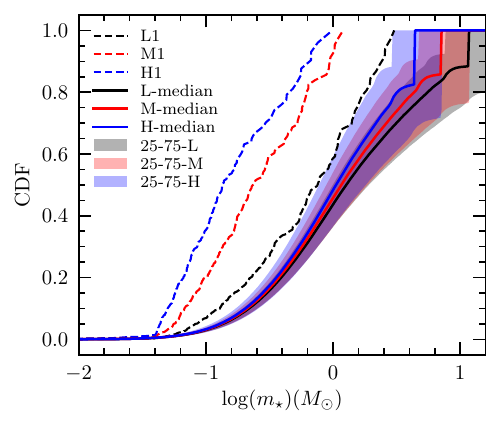}
    \caption{Cumulative distribution function (CDF) of the sink particle masses of the simulations (dashed lines) compared to the CDF for a population of stars with the same total stellar mass as the simulations drawn form a \citet{2005ASSL..327...41C} IMF; for the \citeauthor{2005ASSL..327...41C} IMFs, the central solid line is the 50th percentile value, and the shaded region shows the 25th - 75th percentile range. The black, red and blue lines refer to the low, medium and high surface density runs. Note that the step-like discontinuity at high mass in the \citeauthor{2005ASSL..327...41C} IMF is a real effect, caused by the fact that there is a single most massive star in any real sample, which for the small clusters we are producing contains a not insignificant fraction of the total mass.}
    \label{fig:chabcdf_plot}
\end{figure}

In this section we compare our IMF with the observable IMFs. For the purpose of quantitative comparison between simulation and observation we compare the CDFs  from our simulations with the \citet{2005ASSL..327...41C} IMF. Since our simulations do not produce enough stars to fully sample the high mass end of the mass distribution, it is important to perform this comparison accounting for stochastic effects. For this purpose we generate 50,000 clusters with masses of 100, 50, and 25 $M_{\odot}$, corresponding to the total masses in our L, M, and H runs. To create these clusters we draw stars from a \citeauthor{2005ASSL..327...41C} IMF truncated at 0.01 $M_\odot$ on the lower end and 120 $M_{\odot}$ on the higher end. In this manner we generate 50,000 sample clusters for each of our target masses; we then compute the CDF for each of the samples, from which we can compute the the $p$th percentile value of the CDF at any given stellar mass $m_\star$, which we can compare to our simulation CDFs as shown in \autoref{fig:cdf_plot}. We show this comparison in \autoref{fig:chabcdf_plot}, plotting the 25th, 50th and 75th percentiles (again computed by mass) for the sample clusters.

\autoref{fig:chabcdf_plot} shows the CDFs of runs L1, M1, and H1, along with the CDFs produced by drawing stellar populations of equal mass from a \citeauthor{2005ASSL..327...41C} IMF. It is obvious from the plot that run L1 is reasonably close to a Chabrier IMF, while both runs M1 and H1 are shifted to much lower masses. Much of this shift is at the high mass end, where our simulations do not form as many massive stars as would be expected from an unbiased draw from the IMF. In the L1 runs 40$\%$ of the total stellar mass is stored in stars $>$ 1 $M_{\odot}$ while the H run produces no stars $>$ 1 $M_{\odot}$. This is at least in part a time effect, in the sense that, while we have run long enough for the median mass in our simulations to stabilise, we have not run long enough for the top of the mass distribution to do so; indeed, this is obvious simply from the fact that, since we stop the simulations at 5\% SFE, we could not have formed a star more massive than 50 M$_\odot$ in run M (100 M$_\odot$ in L, 25 M$_\odot$ in H) even if all the collapsed mass had gone into a single object. Accounting for this effect, it is clear that the typical stellar mass in run L1 is reasonably close to what might be expected from a Charbrier IMF, while that in runs M1 and H1 is substantially lower.

%Our results demonstrates that as we go up in the surface density in the simulations, the IMF becomes 
\subsection{Implications for the mass to light ratio in early type galaxies}
\label{ssec:mass_to_light}

As discussed in \autoref{sec:intro}, one of the primary motivations for this work is to attempt to understand the variations in apparent IMF that have been observed in early type galaxies. While there are multiple lines of evidence for this variation, the most straightforward to extract from out simulations is the mass to light ratio of the stellar populations we produce. In this section we therefore look at the mass to light ratio in our simulations for the purpose of comparing to that in observed galaxies. We calculate this by using the \textsc{slug} stellar population synthesis code \citep{da-Silva12a, Krumholz15b} to generate isochrones at stellar population ages from 5 Gyr to 10 Gyr, using the MIST stellar evolution tracks \citep{Choi16a} and Starburst99-style stellar atmosphere models \citep{Leitherer99a}. Each isochrone provides a prediction of present-day mass, bolometric luminosity, and luminosity in a range of photometric filters as a function of initial mass for stars with initial mass $\geq 0.1$ M$_\odot$; we assume that the luminosities of stars with initial masses less than 0.1 M$_{\odot}$ are negligible, and that these stars also experience negligible mass loss. We further assume that all stars with initial mass $<8$ M$_\odot$ (which are all the stars formed in our simulations) that reach the end of their lives leave behind 0.7 M$_\odot$ white dwarf remnants. We use the isochrone to calculate the luminosity and present-day mass of all the stars formed in each of our simulations at ages from $5-10$ Gyr, and from these we calculate the mass to light ratio of the stellar population as a function of age for each of our simulations. For comparison, we use the same isochrones to calculate the mass to light ratio of \citet{2005ASSL..327...41C} and \citet[truncated at a lower mass limit of $0.1$ M$_\odot$]{1955ApJ...121..161S} IMFs at the same ages. We use the SDSS $r$ band for this calculation, but results are qualitatively similar in other filters.% at similar wavelengths.

\begin{figure}
	\includegraphics[width=1.0\columnwidth]{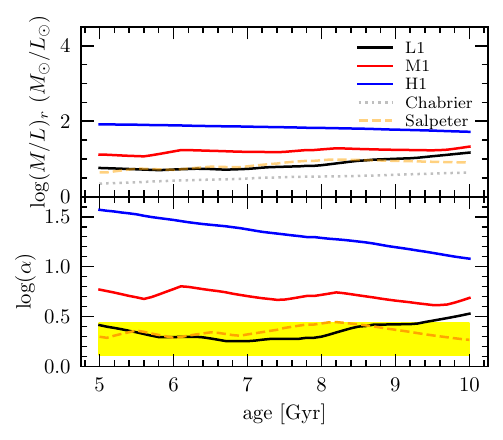}
    \caption{The top panel shows the mass to light ratio of the stellar population formed in the simulations in Solar units, and for comparison the mass to light ratio of stars drawn from a \citet{2005ASSL..327...41C} IMF and a \citet{1955ApJ...121..161S} (dashed lines), all as a function of age. The black, red, blue solid lines are of the simulations L1, M1, H1 respectively. The bottom panel shows the IMF mismatch parameter $\alpha = (M/L)/(M/L)_{\rm Chabrier}$, again as a function of stellar population age. The yellow band shows the range of $\alpha$ values of 41 ETG from \citet{2021arXiv211011985G}}.
    \label{fig:ml_plot}
\end{figure}

% MRK I have held off on commenting on this next paragraph pending the final version of the plot
%From \autoref{fig:ml_plot} shows the mass to light ratio from a Chabrier IMF under predicts the mass to light ratio of all the runs. We can also see from the plot that the \citet{1955ApJ...121..161S} under predicts the mass to light ratio produced by H1 and M1 in this case. Run L1 produces a mass to light ratio that is closer to the mass to light ratio produced by the \citet{1955ApJ...121..161S} IMF. This indicates that the IMF produced by the runs M1, H1 are more bottom heavy than a \citet{1955ApJ...121..161S} IMF. All the runs however are more bottom heavy than the \citet{2005ASSL..327...41C} IMF. 

The upper panel of \autoref{fig:ml_plot} shows the mass to light ratio for the stellar populations produced in our simulations, and for comparison the dashed and dotted lines show the mass to light ratios of populations drawn from \citeauthor{2005ASSL..327...41C} and \citeauthor{1955ApJ...121..161S} IMFs at the same age.  We see that L1 has a mass to light ratio intermediate similar to or slightly lower than Salpeter, M1 is a few tenths of a dex heavier than Salpeter, while H1 is much heavier. Thus the IMF variations we measure in our simulations would manifest as mass to light ratio variations in observations at levels comparable to those observed between early type galaxies and local spirals. The differences between the simulations, and between the simulations and the Chabrier and Salpeter IMF, are much larger than can plausibly be explained by sampling effects associated with the difference in total stellar mass between the runs, as we demonstrate in \aref{app:finite_mass}.

To emphasise the similarity between the $M/L$ variations we find in our simulations and those seen in observations, the lower panel of \autoref{fig:ml_plot} shows the IMF mismatch parameter, $\alpha = (M/L)/(M/L)_{\rm Chabrier}$; i.e., $\alpha=1$ corresponds to a stellar population that has the same $M/L$ as a \citeauthor{2005ASSL..327...41C} IMF. For comparison, the yellow band shows the range of $\alpha$ measured for 41 early type galaxies (ETGs) by \citet{2021arXiv211011985G}; the band shown shows the range that contains 90\% of their sample values. The dashed line shows the mismatch parameter of \citeauthor{1955ApJ...121..161S} IMF. \citeauthor{2021arXiv211011985G} find that most of their sample galaxies have IMFs that deviate from their reference \citeauthor{2005ASSL..327...41C} IMF. In \autoref{fig:ml_plot} we can see that run L1's mismatch parameter is similar to that measured by \citeauthor{2021arXiv211011985G}, while M1 has a slightly heavier IMF and H1 a much heavier one. The absolute value of $\alpha$ is probably not of tremendous significance here, due to the systematic uncertainties in both the simulations and the observations. The more significant point is that our L and M runs are producing differences in mass to light ratio comparable to those observed between ETGs and local spirals. They therefore provide a plausible explanation for how such $M/L$ variations could arise.

\subsection{Effects of systematic variations of SFE}
\label{ssec:effectsofsysSFE}
In this paper, we have run our simulations up to 5$\%$ SFE. At this fixed SFE, the median and the percentile ranges of the stellar mass do approach a steady state. However, the IMF is not fully converged at this point,  because the simulations do not produce enough massive stars to fully sample the IMF at the massive end. This has potentially important implications for variations of the IMF, because \citet{2022MNRAS.515.4929G} find that the final SFE of a cloud varies systematically with the cloud surface density. They argue that this variation suppresses variations in the IMF, because in their simulations the lower stellar masses found in higher surface density clouds at fixed SFE are largely cancelled by a systematic increase in stellar mass with SFE, coupled with higher SFEs in higher surface density clouds.

\begin{figure}
	\includegraphics[width=1.0\columnwidth]{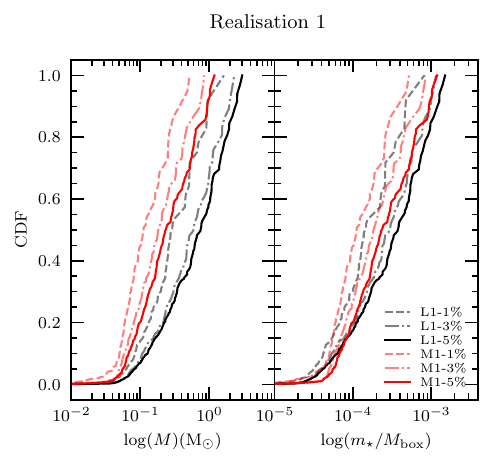}
    \caption{{Same as \autoref{fig:cdf_plot}, but showing CDFs of the sink particle masses of runs L1 and M1 at three different SFEs. The red and black solid lines shows the CDF at 5$\%$ SFE while the dashed and dash-dotted line represents the CDF at 1$\%$ and 3$\%$ SFE.}
    \label{fig:cdfsfe_plot}
    }
\end{figure}

While our simulations do not contain massive stars capable of disrupting clouds and thus limiting the final SFE, even if there were systematic SFE variations, it seems unlikely that this effect could fully remove the IMF variations we have found. To demonstrate this, in \autoref{fig:cdfsfe_plot} we show the mass functions of run L1 and M1 at 1$\%$, 3$\%$, and 5$\%$ SFE; we omit H1 and realisation 2 to avoid clutter, but the results are qualitatively similar. Significantly, we see that (1) the CDFs at 3\% and 5\% are very similar, so replacing one with the other would make relatively little difference, and (2) even if the SFE were to vary much more extremely, for example reaching 1\% in the low surface density case and 5\% in the medium surface density case, the IMF at medium surface density would \textit{still} be more bottom-heavy than at low surface density. This plot clearly shows that while a systematic increase of SFE with surface density would decrease but not completely remove systematic variations of the IMF. 

We also caution against accepting uncritically the conclusion that the final SFE in a real galactic environment will match that predicted by simulations of isolated molecular clouds. These simulations assume that clouds can be disrupted only by internal feedback, and that the external environment plays no role. However, it is far from clear that this is true in realistic galactic environment. For example, \citet{Jeffreson18a} and \citet{Jeffreson20a} find that giant molecular cloud lifetimes are mostly set by internal feedback in low-pressure Milky Way outer disc-like environments, but that at higher pressures, as would be relevant for our M and H runs, environmental effects such as galactic shear, external pressure, and ongoing accretion become increasingly important. It is therefore not clear that the final SFE in a realistic environment will follow the isolated-cloud scaling found in \citet{2022MNRAS.515.4929G}.

\subsection{Alternative scaling}
In our simulations we have kept the dimensionless parameters constant by holding the gas temperature $T$ (or equivalently the sound speed $c_s$) constant while the box mass and box size vary. In the absence of any kind of feedback mechanism we expect the stellar mass to scale just exactly like the Jeans mass, $M_{J} \propto \rho^{-1/2}T^{3/2}$, which, given the relationship between density and surface density in our scalings, implies $M_J \propto \Sigma^{-1}$. Radiation feedback flattens this relationship considerably in our simulations, because radiation is trapped more effectively and thus provides stronger feedback at higher $\Sigma$. This leads to a characteristic stellar mass that still decreases with $\Sigma$, but significantly less strongly than $\Sigma^{-1}$.

However, holding $T$ constant while varying the box size and mass is not the only possible way to hold the dimensionless numbers (Mach number, Alfv\'en Mach number, virial parameter) constant. It is therefore of interest to consider how the stellar mass might vary if instead of having a constant background temperature and sound speed with a variable box mass, we instead varied the sound speed and temperature while keeping a constant mass in the box. Constant box mass implies that the box length $L$ and density $\rho$ vary with surface density as $L\propto \Sigma^{-1/2}$ and $\rho\propto L^{-3}\propto \Sigma^{3/2}$, respectively. Since the Mach number is constant, the velocity dispersion must scale with sound speed as $\sigma_{v} \propto c_{s}$, and thus with temperature as $\sigma_v \propto T^{1/2}$. Similarly, since the box mass and virial ratio are constant, \autoref{eqnvir} implies $\sigma_v \propto L^{-1/2} \propto \Sigma$, and therefore we can also deduce that $T \propto \Sigma^{1/2}$. Combining these scalings, since the Jeans mass $M_{J} \propto T^{3/2}\rho^{-1/2}$, in this scenario in the absence of any feedback mechanism, $M_{J}$ will be independent of $\Sigma$. We can think of this as being due to a cancellation: a smaller cloud of fixed mass has higher density, which favours fragmentation to smaller masses, but also requires higher velocity dispersion to remain at constant virial ratio, which in turn implies higher temperature if the Mach number is also constant. This higher temperature suppresses fragmentation, exactly cancelling the density effect.

As in our fiducial scaling where we hold $T$ constant, this provides a zeroth-order answer to how we might expect the characteristic stellar mass to vary with the surface density of the environment. The first-order answer must then incorporate the effects of feedback, and in particular how radiative suppression of fragmentation will operate as a function of $\Sigma$. The main effect we have identified in this paper will still operate, in that higher $\Sigma$ clouds will trap radiation more effectively, and thus we expect radiation feedback to be more effective at higher $\Sigma$. However, in the alternative scaling scenario where we vary $T$, there is a countervailing effect that we must consider. In a cloud where the background temperature is lower , we expect that radiation feedback will be more effective at pushing the characteristic mass upwards, since injecting the same amount of thermal energy from starlight will lead to a larger perturbation to the temperature. Conversely, radiation feedback will be less effective when the background temperature is higher, resulting in a lower characteristic mass relative to the box mass despite the jeans mass being higher. Since $T\propto \sqrt{\Sigma}$ in this alternative scaling, we might expect radiation feedback to be \textit{less} effective rather than more effective at higher $\Sigma$. It is unclear, and we probably cannot determine absent full simulations, whether this effect is stronger or weaker than the optical depth effect, which points in the opposite direction. Thus it is unclear whether the resulting characteristic stellar mass with feedback would be slightly increasing or slightly decreasing with $\Sigma$; in either case, though, it seems likely to scale with $\Sigma$ less strongly than for our fiducial case.

We emphasise, however, that while this alternative scaling is useful as a thought exercise, it is unlikely to be astrophysically realised. While there are good physical reasons for the temperatures in star-forming clouds to be only weakly dependent on galactic environment (mostly that CO and dust cooling provide very effective thermostats -- e.g., see \citealt{2022MNRAS.509.1959S}), there is no obvious reason why clouds should have fixed Mach numbers. Indeed, observations suggest that they clearly do not, since the observed molecular gas velocity dispersions in some starburst galaxies easily reach ten times those found in the Milky Way \citep[e.g.,][]{Scoville17a}, but the gas temperatures are not 100 times larger, as would be required for constant $\mathcal{M}$. Nonetheless, to the extent that the background temperature does increase with $\Sigma$, even if not enough to maintain constant $\mathcal{M}$, our analysis above suggests that this would somewhat flatten the dependence of stellar mass on $\Sigma$ compared to our numerical results.

\section{Conclusions}
\label{Conclusion}
In this paper we present a new set of magnetohydrodynamic simulations including both radiation and protostellar outflow feedback, which we use to explore environmental variations of the initial mass function. The goal of these simulations is to isolate the interaction of feedback with the star-forming environment, something we achieve by setting up experiments in which we keep the dimensionless parameters constant so that, in the absence of feedback, our simulations representing different star-forming environments would all simply be rescaled versions of one another, and thus would yield statistically identical mass distributions when normalised to the total simulation mass. We then systematically vary the surface density of the star-forming environment, carrying out two different realisations for each environment of interest to obtain more robust statistics.

We find that as the surface density increases the median stellar mass decreases and the mass range of the IMF becomes narrower. Both the radiation feedback and protostellar outflow feedback contribute to these differences, but in different ways. Radiation feedback suppresses fragmentation to smaller objects, and does so with increasing effectiveness as the surface density rises and radiation is trapped more effectively. This pushes the low end of the IMF upwards in higher surface density environments, though not by enough to fully compensate for the decrease in Jeans mass with increasing density.

However, if this were the only feedback mechanism, we would expect the entire IMF to shift upward due to radiation, which is not what we find. Instead, we find that outflow feedback is dominant at the high mass end of the IMF. When the surface density is low, stars contract further towards the main sequence during star formation, yielding more powerful outflows that can easily escape to low density regions where they are able to expand and occupy more volume. Conversely, this makes them less effective as a feedback mechanism, since it leads to less fragmentation of the dense gas. By contrast, weaker outflows are trapped in dense gas, where they break up massive clumps more effectively, suppressing massive star formation. The combination of these two effects -- more effective outflow suppression of high-mass stars at high surface density, and more effective radiative suppression of low mass stars -- together lead to a narrower IMF at higher surface density.

We also explore the implications of our simulations for understanding the mass to light ratio variations observed in early type galaxies.  The IMF variations we find produce shifts in the mass to light ratio of old stellar populations that are qualitatively consistent with those inferred from observations. The shifts are also directionally correct, in the sense that denser star-forming environments, as might be expected to characterise the environments in which the stars of early type galaxies formed, give rise to stellar populations with higher mass to light ratios once they reach $\sim 10$ Gyr ages. Thus our results provide a potential explanation for the origin of these variations.

%with the mass to light ratio produced by a \citet{2005ASSL..327...41C}. We find that the \citet{2005ASSL..327...41C} under predicts the mass to light ratio found in run M1 and H1. This finding is line with what  \citet{2012Natur.484..485C} have found in early type galaxies.

%In our simulations we have found that as the surface density increases the peak of the IMF shifts to much lower masses. This suggest that the IMF may not be universal rather it is dependent on the interaction with the environment and the feedback process. 

\section*{Acknowledgements}

MRK acknowledges support from the Australian Research Council through its Future Fellowship and Discovery Projects funding schemes, awards FT180100375 and DP190101258. CF acknowledges funding provided by the Australian Research Council (Future Fellowship FT180100495), and the Australia-Germany Joint Research Cooperation Scheme (UA-DAAD). This research was undertaken with the assistance of resources and services from the National Computational Infrastructure (NCI, grants jh2 and ek9), which is supported by the Australian Government, awarded through the National Merit, ANU Merit, and Astronomy Supercomputing Time grant schemes.

%%%%%%%%%%%%%%%%%%%%%%%%%%%%%%%%%%%%%%%%%%%%%%%%%%
\section*{Data Availability}

The data underlying this article will be shared on reasonable request to the corresponding author.



\bibliographystyle{mnras}
\bibliography{example,federrath} % if your bibtex file is called example.bib

\begin{thebibliography}{}
\makeatletter
\relax
\def\mn@urlcharsother{\let\do\@makeother \do\$\do\&\do\#\do\^\do\_\do\%\do\~}
\def\mn@doi{\begingroup\mn@urlcharsother \@ifnextchar [ {\mn@doi@}
  {\mn@doi@[]}}
\def\mn@doi@[#1]#2{\def\@tempa{#1}\ifx\@tempa\@empty \href
  {http://dx.doi.org/#2} {doi:#2}\else \href {http://dx.doi.org/#2} {#1}\fi
  \endgroup}
\def\mn@eprint#1#2{\mn@eprint@#1:#2::\@nil}
\def\mn@eprint@arXiv#1{\href {http://arxiv.org/abs/#1} {{\tt arXiv:#1}}}
\def\mn@eprint@dblp#1{\href {http://dblp.uni-trier.de/rec/bibtex/#1.xml}
  {dblp:#1}}
\def\mn@eprint@#1:#2:#3:#4\@nil{\def\@tempa {#1}\def\@tempb {#2}\def\@tempc
  {#3}\ifx \@tempc \@empty \let \@tempc \@tempb \let \@tempb \@tempa \fi \ifx
  \@tempb \@empty \def\@tempb {arXiv}\fi \@ifundefined
  {mn@eprint@\@tempb}{\@tempb:\@tempc}{\expandafter \expandafter \csname
  mn@eprint@\@tempb\endcsname \expandafter{\@tempc}}}

\bibitem[\protect\citeauthoryear{{Bate}}{{Bate}}{2009}]{Bate09a}
{Bate} M.~R.,  2009, \mn@doi [\mnras] {10.1111/j.1365-2966.2008.14165.x}, \href
  {http://adsabs.harvard.edu/abs/2009MNRAS.392.1363B} {392, 1363}

\bibitem[\protect\citeauthoryear{{Bate}}{{Bate}}{2012}]{Bate12a}
{Bate} M.~R.,  2012, \mn@doi [\mnras] {10.1111/j.1365-2966.2011.19955.x}, \href
  {http://adsabs.harvard.edu/abs/2012MNRAS.419.3115B} {419, 3115}

\bibitem[\protect\citeauthoryear{{Bate}}{{Bate}}{2019}]{Bate19a}
{Bate} M.~R.,  2019, \mn@doi [\mnras] {10.1093/mnras/stz103}, \href
  {https://ui.adsabs.harvard.edu/abs/2019MNRAS.484.2341B} {484, 2341}

\bibitem[\protect\citeauthoryear{{Bate} \& {Keto}}{{Bate} \&
  {Keto}}{2015}]{Bate15a}
{Bate} M.~R.,  {Keto} E.~R.,  2015, \mn@doi [\mnras] {10.1093/mnras/stv451},
  \href {http://adsabs.harvard.edu/abs/2015MNRAS.449.2643B} {449, 2643}

\bibitem[\protect\citeauthoryear{{Burgers}}{{Burgers}}{1948}]{Burgers1948}
{Burgers} J.~M.,  1948, Advances in Applied Mechanics, 1, 171

\bibitem[\protect\citeauthoryear{{Cappellari} et~al.,}{{Cappellari}
  et~al.}{2012}]{2012Natur.484..485C}
{Cappellari} M.,  et~al., 2012, \mn@doi [\nat] {10.1038/nature10972}, \href
  {https://ui.adsabs.harvard.edu/abs/2012Natur.484..485C} {484, 485}

\bibitem[\protect\citeauthoryear{{Chabrier}}{{Chabrier}}{2005}]{2005ASSL..327...41C}
{Chabrier} G.,  2005, {The Initial Mass Function: From Salpeter 1955 to 2005}.
p.~41, \mn@doi{10.1007/978-1-4020-3407-7_5}

\bibitem[\protect\citeauthoryear{{Chakrabarti} \& {McKee}}{{Chakrabarti} \&
  {McKee}}{2005}]{Chakrabarti05a}
{Chakrabarti} S.,  {McKee} C.~F.,  2005, \mn@doi [\apj] {10.1086/432659}, \href
  {http://adsabs.harvard.edu/cgi-bin/nph-bib\_query?bibcode=2005ApJ...631..792C\&db\_key=AST}
  {631, 792}

\bibitem[\protect\citeauthoryear{{Choi}, {Dotter}, {Conroy}, {Cantiello},
  {Paxton}  \& {Johnson}}{{Choi} et~al.}{2016}]{Choi16a}
{Choi} J.,  {Dotter} A.,  {Conroy} C.,  {Cantiello} M.,  {Paxton} B.,
  {Johnson} B.~D.,  2016, \mn@doi [\apj] {10.3847/0004-637X/823/2/102}, \href
  {http://adsabs.harvard.edu/abs/2016ApJ...823..102C} {823, 102}

\bibitem[\protect\citeauthoryear{{Conroy}, {van Dokkum}  \&
  {Villaume}}{{Conroy} et~al.}{2017}]{Conroy17a}
{Conroy} C.,  {van Dokkum} P.~G.,   {Villaume} A.,  2017, \mn@doi [\apj]
  {10.3847/1538-4357/aa6190}, \href
  {https://ui.adsabs.harvard.edu/abs/2017ApJ...837..166C} {837, 166}

\bibitem[\protect\citeauthoryear{{Cunningham}, {Klein}, {Krumholz}  \&
  {McKee}}{{Cunningham} et~al.}{2011}]{2011ApJ...740..107C}
{Cunningham} A.~J.,  {Klein} R.~I.,  {Krumholz} M.~R.,   {McKee} C.~F.,  2011,
  \mn@doi [\apj] {10.1088/0004-637X/740/2/107}, \href
  {https://ui.adsabs.harvard.edu/abs/2011ApJ...740..107C} {740, 107}

\bibitem[\protect\citeauthoryear{{Cunningham}, {Krumholz}, {McKee}  \&
  {Klein}}{{Cunningham} et~al.}{2018}]{2018MNRAS.476..771C}
{Cunningham} A.~J.,  {Krumholz} M.~R.,  {McKee} C.~F.,   {Klein} R.~I.,  2018,
  \mn@doi [\mnras] {10.1093/mnras/sty154}, \href
  {https://ui.adsabs.harvard.edu/abs/2018MNRAS.476..771C} {476, 771}

\bibitem[\protect\citeauthoryear{{El-Badry}, {Weisz}  \& {Quataert}}{{El-Badry}
  et~al.}{2017}]{El-Badry17a}
{El-Badry} K.,  {Weisz} D.~R.,   {Quataert} E.,  2017, \mn@doi [\mnras]
  {10.1093/mnras/stx436}, \href
  {https://ui.adsabs.harvard.edu/abs/2017MNRAS.468..319E} {468, 319}

\bibitem[\protect\citeauthoryear{{Elmegreen}, {Klessen}  \&
  {Wilson}}{{Elmegreen} et~al.}{2008}]{2008ApJ...681..365E}
{Elmegreen} B.~G.,  {Klessen} R.~S.,   {Wilson} C.~D.,  2008, \mn@doi [\apj]
  {10.1086/588725}, \href
  {https://ui.adsabs.harvard.edu/abs/2008ApJ...681..365E} {681, 365}

\bibitem[\protect\citeauthoryear{{Federrath} \& {Klessen}}{{Federrath} \&
  {Klessen}}{2012}]{FederrathKlessen2012}
{Federrath} C.,  {Klessen} R.~S.,  2012, \mn@doi [\apj]
  {10.1088/0004-637X/761/2/156}, \href
  {http://adsabs.harvard.edu/abs/2012ApJ...761..156F} {761, 156}

\bibitem[\protect\citeauthoryear{{Federrath}, {Klessen}  \&
  {Schmidt}}{{Federrath} et~al.}{2008}]{FederrathKlessenSchmidt2008}
{Federrath} C.,  {Klessen} R.~S.,   {Schmidt} W.,  2008, \mn@doi [\apjl]
  {10.1086/595280}, \href {http://adsabs.harvard.edu/abs/2008ApJ...688L..79F}
  {688, L79}

\bibitem[\protect\citeauthoryear{{Federrath}, {Roman-Duval}, {Klessen},
  {Schmidt}  \& {Mac Low}}{{Federrath} et~al.}{2010}]{2010A&A...512A..81F}
{Federrath} C.,  {Roman-Duval} J.,  {Klessen} R.~S.,  {Schmidt} W.,   {Mac Low}
  M.~M.,  2010, \mn@doi [\aap] {10.1051/0004-6361/200912437}, \href
  {https://ui.adsabs.harvard.edu/abs/2010A&A...512A..81F} {512, A81}

\bibitem[\protect\citeauthoryear{{Federrath}, {Sur}, {Schleicher}, {Banerjee}
  \& {Klessen}}{{Federrath} et~al.}{2011}]{2011ApJ...731...62F}
{Federrath} C.,  {Sur} S.,  {Schleicher} D. R.~G.,  {Banerjee} R.,   {Klessen}
  R.~S.,  2011, \mn@doi [\apj] {10.1088/0004-637X/731/1/62}, \href
  {https://ui.adsabs.harvard.edu/abs/2011ApJ...731...62F} {731, 62}

\bibitem[\protect\citeauthoryear{{Federrath}, {Schr{\"o}n}, {Banerjee}  \&
  {Klessen}}{{Federrath} et~al.}{2014}]{FederrathEtAl2014}
{Federrath} C.,  {Schr{\"o}n} M.,  {Banerjee} R.,   {Klessen} R.~S.,  2014,
  \mn@doi [\apj] {10.1088/0004-637X/790/2/128}, \href
  {http://adsabs.harvard.edu/abs/2014ApJ...790..128F} {790, 128}

\bibitem[\protect\citeauthoryear{{Federrath}, {Krumholz}  \&
  {Hopkins}}{{Federrath} et~al.}{2017}]{FederrathKrumholzHopkins2017}
{Federrath} C.,  {Krumholz} M.,   {Hopkins} P.~F.,  2017, in Journal of Physics
  Conference Series. p. 012007, \mn@doi{10.1088/1742-6596/837/1/012007}

\bibitem[\protect\citeauthoryear{{Federrath}, {Klessen}, {Iapichino}  \&
  {Beattie}}{{Federrath} et~al.}{2021}]{FederrathEtAl2021}
{Federrath} C.,  {Klessen} R.~S.,  {Iapichino} L.,   {Beattie} J.~R.,  2021,
  \mn@doi [Nature Astronomy] {10.1038/s41550-020-01282-z}, \href
  {https://ui.adsabs.harvard.edu/abs/2021NatAs...5..365F} {5, 365}

\bibitem[\protect\citeauthoryear{{Geha} et~al.,}{{Geha} et~al.}{2013}]{Geha13a}
{Geha} M.,  et~al., 2013, \mn@doi [\apj] {10.1088/0004-637X/771/1/29}, \href
  {http://adsabs.harvard.edu/abs/2013ApJ...771...29G} {771, 29}

\bibitem[\protect\citeauthoryear{{Gennaro} et~al.,}{{Gennaro}
  et~al.}{2018}]{Gennaro18a}
{Gennaro} M.,  et~al., 2018, \mn@doi [\apj] {10.3847/1538-4357/aaa973}, \href
  {https://ui.adsabs.harvard.edu/abs/2018ApJ...855...20G} {855, 20}

\bibitem[\protect\citeauthoryear{{Goldsmith}}{{Goldsmith}}{2001}]{2001ApJ...557..736G}
{Goldsmith} P.~F.,  2001, \mn@doi [\apj] {10.1086/322255}, \href
  {https://ui.adsabs.harvard.edu/abs/2001ApJ...557..736G} {557, 736}

\bibitem[\protect\citeauthoryear{{Gu}, {Greene}, {Newman}, {Kreisch},
  {Quenneville}, {Ma}  \& {Blakeslee}}{{Gu} et~al.}{2022}]{2021arXiv211011985G}
{Gu} M.,  {Greene} J.~E.,  {Newman} A.~B.,  {Kreisch} C.,  {Quenneville} M.~E.,
   {Ma} C.-P.,   {Blakeslee} J.~P.,  2022, \mn@doi [\apj]
  {10.3847/1538-4357/ac69ea}, \href
  {https://ui.adsabs.harvard.edu/abs/2022ApJ...932..103G} {932, 103}

\bibitem[\protect\citeauthoryear{{Guszejnov}, {Krumholz}  \&
  {Hopkins}}{{Guszejnov} et~al.}{2016}]{2016MNRAS.458..673G}
{Guszejnov} D.,  {Krumholz} M.~R.,   {Hopkins} P.~F.,  2016, \mn@doi [\mnras]
  {10.1093/mnras/stw315}, \href
  {https://ui.adsabs.harvard.edu/abs/2016MNRAS.458..673G} {458, 673}

\bibitem[\protect\citeauthoryear{{Guszejnov}, {Hopkins}, {Grudi{\'c}},
  {Krumholz}  \& {Federrath}}{{Guszejnov} et~al.}{2018}]{2018MNRAS.480..182G}
{Guszejnov} D.,  {Hopkins} P.~F.,  {Grudi{\'c}} M.~Y.,  {Krumholz} M.~R.,
  {Federrath} C.,  2018, \mn@doi [\mnras] {10.1093/mnras/sty1847}, \href
  {https://ui.adsabs.harvard.edu/abs/2018MNRAS.480..182G} {480, 182}

\bibitem[\protect\citeauthoryear{{Guszejnov}, {Grudi{\'c}}, {Hopkins}, {Offner}
   \& {Faucher-Gigu{\`e}re}}{{Guszejnov} et~al.}{2020}]{Guszejnov20a}
{Guszejnov} D.,  {Grudi{\'c}} M.~Y.,  {Hopkins} P.~F.,  {Offner} S. S.~R.,
  {Faucher-Gigu{\`e}re} C.-A.,  2020, \mn@doi [\mnras]
  {10.1093/mnras/staa1883}, \href
  {https://ui.adsabs.harvard.edu/abs/2020MNRAS.496.5072G} {496, 5072}

\bibitem[\protect\citeauthoryear{{Guszejnov}, {Grudi{\'c}}, {Offner},
  {Faucher-Gigu{\`e}re}, {Hopkins}  \& {Rosen}}{{Guszejnov}
  et~al.}{2022}]{2022MNRAS.515.4929G}
{Guszejnov} D.,  {Grudi{\'c}} M.~Y.,  {Offner} S. S.~R.,  {Faucher-Gigu{\`e}re}
  C.-A.,  {Hopkins} P.~F.,   {Rosen} A.~L.,  2022, \mn@doi [\mnras]
  {10.1093/mnras/stac2060}, \href
  {https://ui.adsabs.harvard.edu/abs/2022MNRAS.515.4929G} {515, 4929}

\bibitem[\protect\citeauthoryear{{Hansen}, {Klein}, {McKee}  \&
  {Fisher}}{{Hansen} et~al.}{2012}]{2012ApJ...747...22H}
{Hansen} C.~E.,  {Klein} R.~I.,  {McKee} C.~F.,   {Fisher} R.~T.,  2012,
  \mn@doi [\apj] {10.1088/0004-637X/747/1/22}, \href
  {https://ui.adsabs.harvard.edu/abs/2012ApJ...747...22H} {747, 22}

\bibitem[\protect\citeauthoryear{{Haugb{\o}lle}, {Padoan}  \&
  {Nordlund}}{{Haugb{\o}lle} et~al.}{2018}]{Haugbolle18a}
{Haugb{\o}lle} T.,  {Padoan} P.,   {Nordlund} {\r{A}}.,  2018, \mn@doi [\apj]
  {10.3847/1538-4357/aaa432}, \href
  {https://ui.adsabs.harvard.edu/abs/2018ApJ...854...35H} {854, 35}

\bibitem[\protect\citeauthoryear{{Hennebelle} \& {Chabrier}}{{Hennebelle} \&
  {Chabrier}}{2008}]{2008ApJ...684..395H}
{Hennebelle} P.,  {Chabrier} G.,  2008, \mn@doi [\apj] {10.1086/589916}, \href
  {https://ui.adsabs.harvard.edu/abs/2008ApJ...684..395H} {684, 395}

\bibitem[\protect\citeauthoryear{{Hennebelle} \& {Chabrier}}{{Hennebelle} \&
  {Chabrier}}{2009}]{Hennebelle09a}
{Hennebelle} P.,  {Chabrier} G.,  2009, \mn@doi [\apj]
  {10.1088/0004-637X/702/2/1428}, \href
  {http://adsabs.harvard.edu/abs/2009ApJ...702.1428H} {702, 1428}

\bibitem[\protect\citeauthoryear{{Hennebelle}, {Lee}  \&
  {Chabrier}}{{Hennebelle} et~al.}{2019}]{2019ApJ...883..140H}
{Hennebelle} P.,  {Lee} Y.-N.,   {Chabrier} G.,  2019, \mn@doi [\apj]
  {10.3847/1538-4357/ab3d46}, \href
  {https://ui.adsabs.harvard.edu/abs/2019ApJ...883..140H} {883, 140}

\bibitem[\protect\citeauthoryear{{Hopkins}}{{Hopkins}}{2012}]{2012MNRAS.423.2037H}
{Hopkins} P.~F.,  2012, \mn@doi [\mnras] {10.1111/j.1365-2966.2012.20731.x},
  \href {https://ui.adsabs.harvard.edu/abs/2012MNRAS.423.2037H} {423, 2037}

\bibitem[\protect\citeauthoryear{{Hopkins}}{{Hopkins}}{2013}]{Hopkins13a}
{Hopkins} P.~F.,  2013, \mn@doi [\mnras] {10.1093/mnras/sts704}, \href
  {http://adsabs.harvard.edu/abs/2013MNRAS.430.1653H} {430, 1653}

\bibitem[\protect\citeauthoryear{{Jappsen}, {Klessen}, {Larson}, {Li}  \& {Mac
  Low}}{{Jappsen} et~al.}{2005}]{JappsenEtAl2005}
{Jappsen} A.-K.,  {Klessen} R.~S.,  {Larson} R.~B.,  {Li} Y.,   {Mac Low}
  M.-M.,  2005, \mn@doi [\aap] {10.1051/0004-6361:20042178}, \href
  {http://cdsads.u-strasbg.fr/abs/2005A%26A...435..611J} {435, 611}

\bibitem[\protect\citeauthoryear{{Jeffreson} \& {Kruijssen}}{{Jeffreson} \&
  {Kruijssen}}{2018}]{Jeffreson18a}
{Jeffreson} S. M.~R.,  {Kruijssen} J.~M.~D.,  2018, \mn@doi [\mnras]
  {10.1093/mnras/sty594}, \href
  {https://ui.adsabs.harvard.edu/abs/2018MNRAS.476.3688J} {476, 3688}

\bibitem[\protect\citeauthoryear{{Jeffreson}, {Kruijssen}, {Keller}, {Chevance}
   \& {Glover}}{{Jeffreson} et~al.}{2020}]{Jeffreson20a}
{Jeffreson} S. M.~R.,  {Kruijssen} J.~M.~D.,  {Keller} B.~W.,  {Chevance} M.,
  {Glover} S. C.~O.,  2020, \mn@doi [\mnras] {10.1093/mnras/staa2127}, \href
  {https://ui.adsabs.harvard.edu/abs/2020MNRAS.498..385J} {498, 385}

\bibitem[\protect\citeauthoryear{{Klein}, {Fisher}, {McKee}  \&
  {Truelove}}{{Klein} et~al.}{1999}]{1999ASSL..240..131K}
{Klein} R.~I.,  {Fisher} R.~T.,  {McKee} C.~F.,   {Truelove} J.~K.,  1999,
  {Gravitational Collapse and Fragmentation in Molecular Clouds with Adaptive
  Mesh Refinement Hydrodynamics$^{CD}$}.
p.~131, \mn@doi{10.1007/978-94-011-4780-4_44}

\bibitem[\protect\citeauthoryear{{Kolmogorov}}{{Kolmogorov}}{1941}]{Kolmogorov1941c}
{Kolmogorov} A.~N.,  1941, Dokl. Akad. Nauk SSSR, 32, 16

\bibitem[\protect\citeauthoryear{{Konigl} \& {Pudritz}}{{Konigl} \&
  {Pudritz}}{2000}]{Konigl00a}
{Konigl} A.,  {Pudritz} R.~E.,  2000, Protostars and Planets IV, \href
  {http://adsabs.harvard.edu/abs/2000prpl.conf..759K} {pp 759--+}

\bibitem[\protect\citeauthoryear{{Kroupa}}{{Kroupa}}{2001}]{2001MNRAS.322..231K}
{Kroupa} P.,  2001, \mn@doi [\mnras] {10.1046/j.1365-8711.2001.04022.x}, \href
  {https://ui.adsabs.harvard.edu/abs/2001MNRAS.322..231K} {322, 231}

\bibitem[\protect\citeauthoryear{{Krumholz}}{{Krumholz}}{2006}]{2006ApJ...641L..45K}
{Krumholz} M.~R.,  2006, \mn@doi [\apjl] {10.1086/503771}, \href
  {https://ui.adsabs.harvard.edu/abs/2006ApJ...641L..45K} {641, L45}

\bibitem[\protect\citeauthoryear{{Krumholz}}{{Krumholz}}{2011}]{Krumholz11e}
{Krumholz} M.~R.,  2011, \mn@doi [\apj] {10.1088/0004-637X/743/2/110}, \href
  {http://adsabs.harvard.edu/abs/2011ApJ...743..110K} {743, 110}

\bibitem[\protect\citeauthoryear{{Krumholz}}{{Krumholz}}{2014}]{2014PhR...539...49K}
{Krumholz} M.~R.,  2014, \mn@doi [\physrep] {10.1016/j.physrep.2014.02.001},
  \href {https://ui.adsabs.harvard.edu/abs/2014PhR...539...49K} {539, 49}

\bibitem[\protect\citeauthoryear{{Krumholz} \& {McKee}}{{Krumholz} \&
  {McKee}}{2008}]{2008Natur.451.1082K}
{Krumholz} M.~R.,  {McKee} C.~F.,  2008, \mn@doi [\nat] {10.1038/nature06620},
  \href {https://ui.adsabs.harvard.edu/abs/2008Natur.451.1082K} {451, 1082}

\bibitem[\protect\citeauthoryear{{Krumholz}, {McKee}  \& {Klein}}{{Krumholz}
  et~al.}{2004}]{2004ApJ...611..399K}
{Krumholz} M.~R.,  {McKee} C.~F.,   {Klein} R.~I.,  2004, \mn@doi [\apj]
  {10.1086/421935}, \href
  {https://ui.adsabs.harvard.edu/abs/2004ApJ...611..399K} {611, 399}

\bibitem[\protect\citeauthoryear{{Krumholz}, {Klein}  \& {McKee}}{{Krumholz}
  et~al.}{2007}]{2007ApJ...656..959K}
{Krumholz} M.~R.,  {Klein} R.~I.,   {McKee} C.~F.,  2007, \mn@doi [\apj]
  {10.1086/510664}, \href
  {https://ui.adsabs.harvard.edu/abs/2007ApJ...656..959K} {656, 959}

\bibitem[\protect\citeauthoryear{{Krumholz}, {Cunningham}, {Klein}  \&
  {McKee}}{{Krumholz} et~al.}{2010}]{2010ApJ...713.1120K}
{Krumholz} M.~R.,  {Cunningham} A.~J.,  {Klein} R.~I.,   {McKee} C.~F.,  2010,
  \mn@doi [\apj] {10.1088/0004-637X/713/2/1120}, \href
  {https://ui.adsabs.harvard.edu/abs/2010ApJ...713.1120K} {713, 1120}

\bibitem[\protect\citeauthoryear{{Krumholz}, {Klein}  \& {McKee}}{{Krumholz}
  et~al.}{2012}]{Krumholz12b}
{Krumholz} M.~R.,  {Klein} R.~I.,   {McKee} C.~F.,  2012, \mn@doi [\apj]
  {10.1088/0004-637X/754/1/71}, \href
  {http://adsabs.harvard.edu/abs/2012ApJ...754...71K} {754, 71}

\bibitem[\protect\citeauthoryear{{Krumholz}, {Fumagalli}, {da Silva}, {Rendahl}
   \& {Parra}}{{Krumholz} et~al.}{2015}]{Krumholz15b}
{Krumholz} M.~R.,  {Fumagalli} M.,  {da Silva} R.~L.,  {Rendahl} T.,   {Parra}
  J.,  2015, \mn@doi [\mnras] {10.1093/mnras/stv1374}, \href
  {http://adsabs.harvard.edu/abs/2015MNRAS.452.1447K} {452, 1447}

\bibitem[\protect\citeauthoryear{{Krumholz}, {Myers}, {Klein}  \&
  {McKee}}{{Krumholz} et~al.}{2016}]{Krumholz16c}
{Krumholz} M.~R.,  {Myers} A.~T.,  {Klein} R.~I.,   {McKee} C.~F.,  2016,
  \mn@doi [\mnras] {10.1093/mnras/stw1236}, \href
  {http://adsabs.harvard.edu/abs/2016MNRAS.460.3272K} {460, 3272}

\bibitem[\protect\citeauthoryear{{La Barbera}, {Ferreras}, {Vazdekis}, {de la
  Rosa}, {de Carvalho}, {Trevisan}, {Falc{\'o}n-Barroso}  \&
  {Ricciardelli}}{{La Barbera} et~al.}{2013}]{La-Barbera13a}
{La Barbera} F.,  {Ferreras} I.,  {Vazdekis} A.,  {de la Rosa} I.~G.,  {de
  Carvalho} R.~R.,  {Trevisan} M.,  {Falc{\'o}n-Barroso} J.,   {Ricciardelli}
  E.,  2013, \mn@doi [\mnras] {10.1093/mnras/stt943}, \href
  {https://ui.adsabs.harvard.edu/abs/2013MNRAS.433.3017L} {433, 3017}

\bibitem[\protect\citeauthoryear{{Larson}}{{Larson}}{2005}]{2005MNRAS.359..211L}
{Larson} R.~B.,  2005, \mn@doi [\mnras] {10.1111/j.1365-2966.2005.08881.x},
  \href {https://ui.adsabs.harvard.edu/abs/2005MNRAS.359..211L} {359, 211}

\bibitem[\protect\citeauthoryear{{Lee} \& {Hennebelle}}{{Lee} \&
  {Hennebelle}}{2018}]{2018A&A...611A..89L}
{Lee} Y.-N.,  {Hennebelle} P.,  2018, \mn@doi [\aap]
  {10.1051/0004-6361/201731523}, \href
  {https://ui.adsabs.harvard.edu/abs/2018A&A...611A..89L} {611, A89}

\bibitem[\protect\citeauthoryear{{Lee}, {Cunningham}, {McKee}  \&
  {Klein}}{{Lee} et~al.}{2014}]{2014ApJ...783...50L}
{Lee} A.~T.,  {Cunningham} A.~J.,  {McKee} C.~F.,   {Klein} R.~I.,  2014,
  \mn@doi [\apj] {10.1088/0004-637X/783/1/50}, \href
  {https://ui.adsabs.harvard.edu/abs/2014ApJ...783...50L} {783, 50}

\bibitem[\protect\citeauthoryear{{Lee}, {Offner}, {Hennebelle}, {Andr{\'e}},
  {Zinnecker}, {Ballesteros-Paredes}, {Inutsuka}  \& {Kruijssen}}{{Lee}
  et~al.}{2020}]{Lee20b}
{Lee} Y.-N.,  {Offner} S. S.~R.,  {Hennebelle} P.,  {Andr{\'e}} P.,
  {Zinnecker} H.,  {Ballesteros-Paredes} J.,  {Inutsuka} S.-i.,   {Kruijssen}
  J.~M.~D.,  2020, \mn@doi [\ssr] {10.1007/s11214-020-00699-2}, \href
  {https://ui.adsabs.harvard.edu/abs/2020SSRv..216...70L} {216, 70}

\bibitem[\protect\citeauthoryear{{Leitherer} et~al.,}{{Leitherer}
  et~al.}{1999}]{Leitherer99a}
{Leitherer} C.,  et~al., 1999, \mn@doi [\apjs] {10.1086/313233}, \href
  {http://adsabs.harvard.edu/cgi-bin/nph-bib\_query?bibcode=1999ApJS..123....3L\&db\_key=AST}
  {123, 3}

\bibitem[\protect\citeauthoryear{{Leroy} et~al.,}{{Leroy}
  et~al.}{2018}]{Leroy18a}
{Leroy} A.~K.,  et~al., 2018, \apj, \href
  {http://adsabs.harvard.edu/abs/2018arXiv180402083L} {}

\bibitem[\protect\citeauthoryear{{Li}, {Martin}, {Klein}  \& {McKee}}{{Li}
  et~al.}{2012}]{2012ApJ...745..139L}
{Li} P.~S.,  {Martin} D.~F.,  {Klein} R.~I.,   {McKee} C.~F.,  2012, \mn@doi
  [\apj] {10.1088/0004-637X/745/2/139}, \href
  {https://ui.adsabs.harvard.edu/abs/2012ApJ...745..139L} {745, 139}

\bibitem[\protect\citeauthoryear{{Li} et~al.,}{{Li} et~al.}{2021}]{Li21b}
{Li} P.,  et~al., 2021, \mn@doi [The Journal of Open Source Software]
  {10.21105/joss.03771}, \href
  {https://ui.adsabs.harvard.edu/abs/2021JOSS....6.3771L} {6, 3771}

\bibitem[\protect\citeauthoryear{{Low} \& {Lynden-Bell}}{{Low} \&
  {Lynden-Bell}}{1976}]{Low76a}
{Low} C.,  {Lynden-Bell} D.,  1976, \mnras, \href
  {http://adsabs.harvard.edu/abs/1976MNRAS.176..367L} {176, 367}

\bibitem[\protect\citeauthoryear{{Mac Low}}{{Mac
  Low}}{1999}]{1999ApJ...524..169M}
{Mac Low} M.-M.,  1999, \mn@doi [\apj] {10.1086/307784}, \href
  {https://ui.adsabs.harvard.edu/abs/1999ApJ...524..169M} {524, 169}

\bibitem[\protect\citeauthoryear{{Masunaga}, {Miyama}  \&
  {Inutsuka}}{{Masunaga} et~al.}{1998}]{1998ApJ...495..346M}
{Masunaga} H.,  {Miyama} S.~M.,   {Inutsuka} S.-i.,  1998, \mn@doi [\apj]
  {10.1086/305281}, \href
  {https://ui.adsabs.harvard.edu/abs/1998ApJ...495..346M} {495, 346}

\bibitem[\protect\citeauthoryear{{Mathew} \& {Federrath}}{{Mathew} \&
  {Federrath}}{2020}]{2020MNRAS.496.5201M}
{Mathew} S.~S.,  {Federrath} C.,  2020, \mn@doi [\mnras]
  {10.1093/mnras/staa1931}, \href
  {https://ui.adsabs.harvard.edu/abs/2020MNRAS.496.5201M} {496, 5201}

\bibitem[\protect\citeauthoryear{{Mathew} \& {Federrath}}{{Mathew} \&
  {Federrath}}{2021}]{2021MNRAS.507.2448M}
{Mathew} S.~S.,  {Federrath} C.,  2021, \mn@doi [\mnras]
  {10.1093/mnras/stab2338}, \href
  {https://ui.adsabs.harvard.edu/abs/2021MNRAS.507.2448M} {507, 2448}

\bibitem[\protect\citeauthoryear{{Miller} \& {Scalo}}{{Miller} \&
  {Scalo}}{1979}]{1979ApJS...41..513M}
{Miller} G.~E.,  {Scalo} J.~M.,  1979, \mn@doi [\apjs] {10.1086/190629}, \href
  {https://ui.adsabs.harvard.edu/abs/1979ApJS...41..513M} {41, 513}

\bibitem[\protect\citeauthoryear{{Myers}, {Krumholz}, {Klein}  \&
  {McKee}}{{Myers} et~al.}{2011}]{Myers11a}
{Myers} A.~T.,  {Krumholz} M.~R.,  {Klein} R.~I.,   {McKee} C.~F.,  2011,
  \mn@doi [\apj] {10.1088/0004-637X/735/1/49}, \href
  {http://adsabs.harvard.edu/abs/2011ApJ...735...49M} {735, 49}

\bibitem[\protect\citeauthoryear{{Myers}, {Klein}, {Krumholz}  \&
  {McKee}}{{Myers} et~al.}{2014}]{2014MNRAS.439.3420M}
{Myers} A.~T.,  {Klein} R.~I.,  {Krumholz} M.~R.,   {McKee} C.~F.,  2014,
  \mn@doi [\mnras] {10.1093/mnras/stu190}, \href
  {https://ui.adsabs.harvard.edu/abs/2014MNRAS.439.3420M} {439, 3420}

\bibitem[\protect\citeauthoryear{{Nam}, {Federrath}  \& {Krumholz}}{{Nam}
  et~al.}{2021}]{Nam21a}
{Nam} D.~G.,  {Federrath} C.,   {Krumholz} M.~R.,  2021, \mn@doi [\mnras]
  {10.1093/mnras/stab505}, \href
  {https://ui.adsabs.harvard.edu/abs/2021MNRAS.503.1138N} {503, 1138}

\bibitem[\protect\citeauthoryear{{Newman}, {Smith}, {Conroy}, {Villaume}  \&
  {van Dokkum}}{{Newman} et~al.}{2017}]{Newman17a}
{Newman} A.~B.,  {Smith} R.~J.,  {Conroy} C.,  {Villaume} A.,   {van Dokkum}
  P.,  2017, \mn@doi [\apj] {10.3847/1538-4357/aa816d}, \href
  {https://ui.adsabs.harvard.edu/abs/2017ApJ...845..157N} {845, 157}

\bibitem[\protect\citeauthoryear{{Offner}, {Klein}, {McKee}  \&
  {Krumholz}}{{Offner} et~al.}{2009}]{2009ApJ...703..131O}
{Offner} S. S.~R.,  {Klein} R.~I.,  {McKee} C.~F.,   {Krumholz} M.~R.,  2009,
  \mn@doi [\apj] {10.1088/0004-637X/703/1/131}, \href
  {https://ui.adsabs.harvard.edu/abs/2009ApJ...703..131O} {703, 131}

\bibitem[\protect\citeauthoryear{{Offner}, {Clark}, {Hennebelle}, {Bastian},
  {Bate}, {Hopkins}, {Moraux}  \& {Whitworth}}{{Offner}
  et~al.}{2014}]{2014prpl.conf...53O}
{Offner} S.~S.~R.,  {Clark} P.~C.,  {Hennebelle} P.,  {Bastian} N.,  {Bate}
  M.~R.,  {Hopkins} P.~F.,  {Moraux} E.,   {Whitworth} A.~P.,  2014, in
  {Beuther} H.,  {Klessen} R.~S.,  {Dullemond} C.~P.,   {Henning} T.,  eds,
  Protostars and Planets VI. p.~53 (\mn@eprint {arXiv} {1312.5326}),
  \mn@doi{10.2458/azu_uapress_9780816531240-ch003}

\bibitem[\protect\citeauthoryear{{Oldham} \& {Auger}}{{Oldham} \&
  {Auger}}{2018}]{Oldham18a}
{Oldham} L.,  {Auger} M.,  2018, \mn@doi [\mnras] {10.1093/mnras/stx2969},
  \href {https://ui.adsabs.harvard.edu/abs/2018MNRAS.474.4169O} {474, 4169}

\bibitem[\protect\citeauthoryear{{Padoan} \& {Nordlund}}{{Padoan} \&
  {Nordlund}}{2002}]{2002ApJ...576..870P}
{Padoan} P.,  {Nordlund} {\r{A}}.,  2002, \mn@doi [\apj] {10.1086/341790},
  \href {https://ui.adsabs.harvard.edu/abs/2002ApJ...576..870P} {576, 870}

\bibitem[\protect\citeauthoryear{{Padoan}, {Nordlund}  \& {Jones}}{{Padoan}
  et~al.}{1997}]{1997MNRAS.288..145P}
{Padoan} P.,  {Nordlund} A.,   {Jones} B. J.~T.,  1997, \mn@doi [\mnras]
  {10.1093/mnras/288.1.145}, \href
  {https://ui.adsabs.harvard.edu/abs/1997MNRAS.288..145P} {288, 145}

\bibitem[\protect\citeauthoryear{{Press} \& {Schechter}}{{Press} \&
  {Schechter}}{1974}]{1974ApJ...187..425P}
{Press} W.~H.,  {Schechter} P.,  1974, \mn@doi [\apj] {10.1086/152650}, \href
  {https://ui.adsabs.harvard.edu/abs/1974ApJ...187..425P} {187, 425}

\bibitem[\protect\citeauthoryear{{Rees}}{{Rees}}{1976}]{Rees76a}
{Rees} M.~J.,  1976, \mnras, \href
  {http://adsabs.harvard.edu/abs/1976MNRAS.176..483R} {176, 483}

\bibitem[\protect\citeauthoryear{{Richer}, {Shepherd}, {Cabrit}, {Bachiller}
  \& {Churchwell}}{{Richer} et~al.}{2000}]{2000prpl.conf..867R}
{Richer} J.~S.,  {Shepherd} D.~S.,  {Cabrit} S.,  {Bachiller} R.,
  {Churchwell} E.,  2000, in {Mannings} V.,  {Boss} A.~P.,   {Russell} S.~S.,
  eds, Protostars and Planets IV. p.~867 (\mn@eprint {arXiv}
  {astro-ph/9904097})

\bibitem[\protect\citeauthoryear{{Salpeter}}{{Salpeter}}{1955}]{1955ApJ...121..161S}
{Salpeter} E.~E.,  1955, \mn@doi [\apj] {10.1086/145971}, \href
  {https://ui.adsabs.harvard.edu/abs/1955ApJ...121..161S} {121, 161}

\bibitem[\protect\citeauthoryear{{Scoville} et~al.,}{{Scoville}
  et~al.}{2017}]{Scoville17a}
{Scoville} N.,  et~al., 2017, \mn@doi [\apj] {10.3847/1538-4357/836/1/66},
  \href {http://adsabs.harvard.edu/abs/2017ApJ...836...66S} {836, 66}

\bibitem[\protect\citeauthoryear{{Semenov}, {Henning}, {Helling}, {Ilgner}  \&
  {Sedlmayr}}{{Semenov} et~al.}{2003}]{Semenov03a}
{Semenov} D.,  {Henning} T.,  {Helling} C.,  {Ilgner} M.,   {Sedlmayr} E.,
  2003, \mn@doi [\aap] {10.1051/0004-6361:20031279}, \href
  {http://adsabs.harvard.edu/abs/2003A\%26A...410..611S} {410, 611}

\bibitem[\protect\citeauthoryear{{Sharda} \& {Krumholz}}{{Sharda} \&
  {Krumholz}}{2022}]{2022MNRAS.509.1959S}
{Sharda} P.,  {Krumholz} M.~R.,  2022, \mn@doi [\mnras]
  {10.1093/mnras/stab2921}, \href
  {https://ui.adsabs.harvard.edu/abs/2022MNRAS.509.1959S} {509, 1959}

\bibitem[\protect\citeauthoryear{{Shu}, {Li}  \& {Allen}}{{Shu}
  et~al.}{2004}]{2004ApJ...601..930S}
{Shu} F.~H.,  {Li} Z.-Y.,   {Allen} A.,  2004, \mn@doi [\apj] {10.1086/380602},
  \href {https://ui.adsabs.harvard.edu/abs/2004ApJ...601..930S} {601, 930}

\bibitem[\protect\citeauthoryear{{Smith}}{{Smith}}{2014}]{Smith14b}
{Smith} R.~J.,  2014, \mn@doi [\mnras] {10.1093/mnrasl/slu082}, \href
  {https://ui.adsabs.harvard.edu/abs/2014MNRAS.443L..69S} {443, L69}

\bibitem[\protect\citeauthoryear{{Smith}}{{Smith}}{2020}]{Smith20a}
{Smith} R.~J.,  2020, \mn@doi [\araa] {10.1146/annurev-astro-032620-020217},
  \href {https://ui.adsabs.harvard.edu/abs/2020ARA&A..58..577S} {58, 577}

\bibitem[\protect\citeauthoryear{{Spaans} \& {Silk}}{{Spaans} \&
  {Silk}}{2000}]{2000ApJ...538..115S}
{Spaans} M.,  {Silk} J.,  2000, \mn@doi [\apj] {10.1086/309118}, \href
  {https://ui.adsabs.harvard.edu/abs/2000ApJ...538..115S} {538, 115}

\bibitem[\protect\citeauthoryear{{Spiniello}, {Trager}, {Koopmans}  \&
  {Chen}}{{Spiniello} et~al.}{2012}]{Spiniello12a}
{Spiniello} C.,  {Trager} S.~C.,  {Koopmans} L.~V.~E.,   {Chen} Y.~P.,  2012,
  \mn@doi [\apjl] {10.1088/2041-8205/753/2/L32}, \href
  {http://adsabs.harvard.edu/abs/2012ApJ...753L..32S} {753, L32}

\bibitem[\protect\citeauthoryear{{Spiniello}, {Koopmans}, {Trager},
  {Barnab{\`e}}, {Treu}, {Czoske}, {Vegetti}  \& {Bolton}}{{Spiniello}
  et~al.}{2015}]{Spiniello15a}
{Spiniello} C.,  {Koopmans} L.~V.~E.,  {Trager} S.~C.,  {Barnab{\`e}} M.,
  {Treu} T.,  {Czoske} O.,  {Vegetti} S.,   {Bolton} A.,  2015, \mn@doi
  [\mnras] {10.1093/mnras/stv1490}, \href
  {http://adsabs.harvard.edu/abs/2015MNRAS.452.2434S} {452, 2434}

\bibitem[\protect\citeauthoryear{{Treu}, {Auger}, {Koopmans}, {Gavazzi},
  {Marshall}  \& {Bolton}}{{Treu} et~al.}{2010}]{Treu10a}
{Treu} T.,  {Auger} M.~W.,  {Koopmans} L. V.~E.,  {Gavazzi} R.,  {Marshall}
  P.~J.,   {Bolton} A.~S.,  2010, \mn@doi [\apj]
  {10.1088/0004-637X/709/2/1195}, \href
  {https://ui.adsabs.harvard.edu/abs/2010ApJ...709.1195T} {709, 1195}

\bibitem[\protect\citeauthoryear{{Truelove}, {Klein}, {McKee}, {Holliman},
  {Howell}  \& {Greenough}}{{Truelove} et~al.}{1997}]{1997ApJ...489L.179T}
{Truelove} J.~K.,  {Klein} R.~I.,  {McKee} C.~F.,  {Holliman} John~H. I.,
  {Howell} L.~H.,   {Greenough} J.~A.,  1997, \mn@doi [\apjl] {10.1086/310975},
  \href {https://ui.adsabs.harvard.edu/abs/1997ApJ...489L.179T} {489, L179}

\bibitem[\protect\citeauthoryear{{Truelove}, {Klein}, {McKee}, {Holliman},
  {Howell}, {Greenough}  \& {Woods}}{{Truelove}
  et~al.}{1998}]{1998ApJ...495..821T}
{Truelove} J.~K.,  {Klein} R.~I.,  {McKee} C.~F.,  {Holliman} John~H. I.,
  {Howell} L.~H.,  {Greenough} J.~A.,   {Woods} D.~T.,  1998, \mn@doi [\apj]
  {10.1086/305329}, \href
  {https://ui.adsabs.harvard.edu/abs/1998ApJ...495..821T} {495, 821}

\bibitem[\protect\citeauthoryear{{Urquhart} et~al.,}{{Urquhart}
  et~al.}{2018}]{Urquhart18a}
{Urquhart} J.~S.,  et~al., 2018, \mn@doi [\mnras] {10.1093/mnras/stx2258},
  \href {https://ui.adsabs.harvard.edu/#abs/2018MNRAS.473.1059U} {473, 1059}

\bibitem[\protect\citeauthoryear{{da Silva}, {Fumagalli}  \& {Krumholz}}{{da
  Silva} et~al.}{2012}]{da-Silva12a}
{da Silva} R.~L.,  {Fumagalli} M.,   {Krumholz} M.,  2012, \mn@doi [\apj]
  {10.1088/0004-637X/745/2/145}, \href
  {http://adsabs.harvard.edu/abs/2012ApJ...745..145D} {745, 145}

\bibitem[\protect\citeauthoryear{{van Dokkum} \& {Conroy}}{{van Dokkum} \&
  {Conroy}}{2010}]{2010Natur.468..940V}
{van Dokkum} P.~G.,  {Conroy} C.,  2010, \mn@doi [\nat] {10.1038/nature09578},
  \href {https://ui.adsabs.harvard.edu/abs/2010Natur.468..940V} {468, 940}

\makeatother
\end{thebibliography}

%\begin{thebibliography}{99}
%\end{thebibliography}

%%%%%%%%%%%%%%%%%%%%%%%%%%%%%%%%%%%%%%%%%%%%%%%%%%

%%%%%%%%%%%%%%%%% APPENDICES %%%%%%%%%%%%%%%%%%%%%

\appendix
\section{Mass to light ratios with finite masses} 
\label{app:finite_mass}

Our simulations differ from each other in the total mass of stars they produce, and the finite masses found in them also differ from the infinitely-sampled limit we use to calculate our fiducial mass to light ratios for the \citet{1955ApJ...121..161S} and \citet{2005ASSL..327...41C} IMFs. It is therefore important to verify that the differences in mass to light ratio that we find are not due to the effects of sampling different total stellar masses in each case. To carry out this check, we repeat the \textsc{slug} calculations described in \autoref{ssec:mass_to_light} to create synthetic clusters with total masses 100, 50 and 25 M$_{\odot}$, corresponding to the total stellar masses in L, M and H runs. For this experiment we draw stars from a \citet{2005ASSL..327...41C} IMF truncated at 0.01 $M_{\odot}$ at the low mass end and 120 $M_{\odot}$ at the high mass end. We generate 50,000 clusters for each of our target masses, and for each run we compute $M/L$ as a function of age from 5 - 10 Gyr, exactly as in \autoref{ssec:mass_to_light}.

\begin{figure}
	\includegraphics[width=1.0\columnwidth]{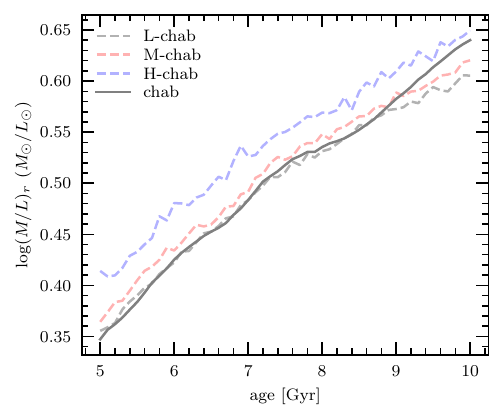}
    \caption{The solid black line shows the mass to light ratio of an infinitely-massive simple stellar population with a \citet{2005ASSL..327...41C} IMF as a function of age. The black, red, and blue dashed lines show the mean mass to light ratios of stellar populations drawn from the same IMF with total masses of 100, 50, and 25 M$_\odot$, corresponding to the total stellar masses in runs L, M, and H.}
    \label{fig:darth_plot}
\end{figure}

\autoref{fig:darth_plot} shows the mean mass to light ratios of these finite-mass experiments, in comparison to the infinitely-sampled limit. We see that, not surprisingly, the effects are largest for run H, which has the smallest total mass, but that the effects overall are never large. The maximum difference between the finite-mass and infinitely-sampled results never exceeds $\approx 0.05$ dex. Comparison to \autoref{fig:ml_plot} shows that such differences are much smaller than the differences we find between the different runs, and between them and the \citet{2005ASSL..327...41C} IMF.

%%%%%%%%%%%%%%%%%%%%%%%%%%%%%%%%%%%%%%%%%%%%%%%%%%

% Don't change these lines
\bsp	% typesetting comment
\label{lastpage}
\end{document}